\documentclass[twocolumn,preprintnumbers,amsmath,amssymb,superscriptaddress]{revtex4-1}
\usepackage{graphicx}
\usepackage{subcaption}
\usepackage{amsmath}
\usepackage{subcaption}
\usepackage{algorithm}
\usepackage{algorithmic}
\usepackage{color}
\usepackage{ulem}

\newcommand{\commentoutA}[1]{}

\begin{document}

\preprint{LA-UR-23-22319}
\preprint{SAND2023-01051R}

\title{Shadow molecular dynamics and atomic cluster expansions for flexible charge models}

\author{James Goff $^{\, *}$}
\affiliation{Center for Computing and Research, Sandia National Laboratories, Albuquerque, New Mexico 871 85, USA}
\email{jmgoff@sandia.gov}
\author{Yu Zhang}
\affiliation{Theoretical Division, Los Alamos National Laboratory, Los Alamos, New Mexico 87545, USA}
\author{Christian Negre}
\affiliation{Theoretical Division, Los Alamos National Laboratory, Los Alamos, New Mexico 87545, USA}
\author{Andrew Rohskopf}
\affiliation{Center for Computing and Research, Sandia National Laboratories, Albuquerque, New Mexico 871 85, USA}
\author{Anders M. N. Niklasson}
\email{amn@lanl.gov}
\affiliation{Theoretical Division, Los Alamos National Laboratory, Los Alamos, New Mexico 87545, USA}

\date{\today}

\begin{abstract}
A shadow molecular dynamics scheme for flexible charge models is presented, where the shadow Born-Oppenheimer potential is derived from a coarse-grained approximation of range-separated density functional theory. The interatomic potential, including the atomic electronegativities and the charge-independent short-range part of the potential and force terms, are modeled by the linear atomic cluster expansion (ACE), which provides a computationally efficient alternative to many machine learning methods.
The shadow molecular dynamics scheme is based on extended Lagrangian (XL) Born-Oppenheimer molecular dynamics (BOMD) [Eur.\ Phys.\ J.\ B {\bf 94}, 164 (2021)]. XL-BOMD provides a stable dynamics, while avoiding the costly computational overhead associated with solving an all-to-all system of equations, which normally is required to determine the relaxed electronic ground state prior to each force evaluation. To demonstrate the proposed shadow molecular dynamics scheme for flexible charge models using the atomic cluster expansion, we emulate the dynamics generated from self-consistent charge density functional tight-binding (SCC-DFTB) theory using a second-order charge equilibration (QEq) model. The charge-independent potentials and electronegativities of the QEq model are trained for a supercell of uranium oxide (UO$_2$) and a molecular system of liquid water. The combined ACE + XL-QEq molecular dynamics simulations are stable over a wide range of temperatures both for the oxide and the molecular systems, and provide a precise sampling of the Born-Oppenheimer potential energy surfaces. Accurate ground Coulomb energies are produced by the ACE-based electronegativity model during an NVE simulation of UO$_2$, predicted to be within 1 meV of those from SCC-DFTB on average during comparable simulations. 
\end{abstract}

\maketitle
\begin{figure}
    \includegraphics[width=0.95\textwidth]{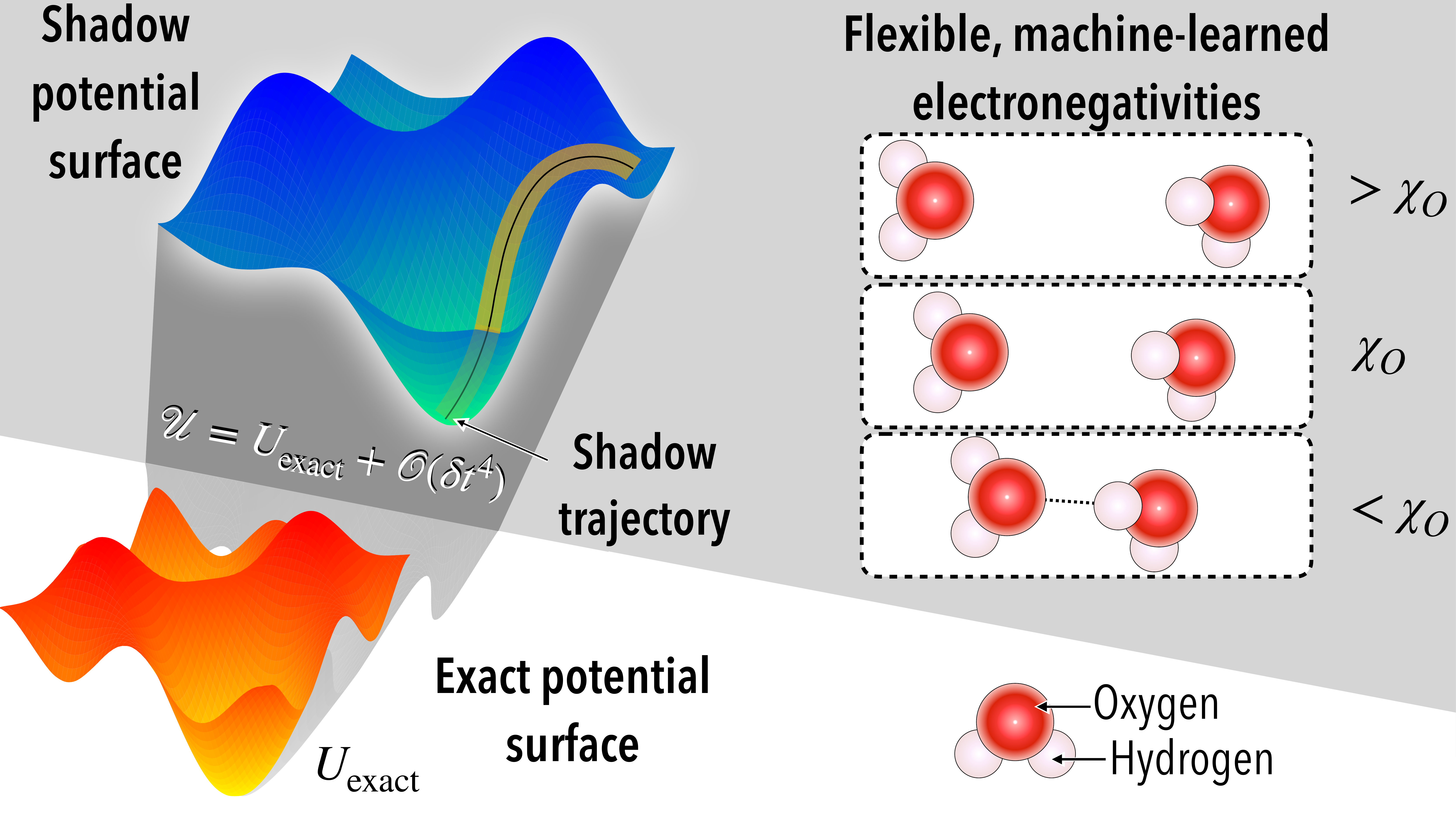}
\end{figure}
\clearpage

\section{Introduction}
Atomistic simulation methods are currently undergoing a revolution thanks to new machine-learning techniques that can provide highly accurate short-range interatomic potentials \cite{JBehler07,APBartok10,MRupp12,APThompson15,Anatole15,JBehler16,NLubbers18,FAFaber17,JHan18,JSSmith19,Drautz19,TMiller20,Tkatchenko20,Drautz21,AClark21}. However, these short-range potentials have limitations, as they do not capture important non-local quantum-mechanical effects, as well as the long-range electrostatic interactions and associated charge relaxations. As a result, they may not be suitable for many real-world applications.

It is possible to account for the long-range charge-dependent interactions and generate more general and accurate atomistic models by combining machine learning techniques with some explicit representation of the electronic structure, for example, polarizable flexible charge models \cite{WJMortier86,AKRappe91,Baker15,JZhifeng19,SGoedecker15,ANiklasson21b,TWKo20,Staacke_2022,Shao_2022}. 
Unfortunately, these methods introduce a dramatic increase in the computational cost, because of the long-range Coulomb interactions between the atom-centered flexible charges, which limits their applicability. The computational cost comes from having to solve all-to-all systems of equations to determine the long-range charge relaxations (or equilibration) prior to each force evaluation as the atoms move. We will present a general formulation for hybrid machine learned and flexible charge models that avoids this overhead.
The theory is based on extended Lagrangian (XL) Born-Oppenheimer molecular dynamics (BOMD) 
\cite{ANiklasson07,ANiklasson08,MCawkwell12,JHutter12,LLin14,PSouvatzis14,KNomura15,AAlbaugh15,ANiklasson17,AAlbaugh18,ANiklasson21b} in its most recent \textit{shadow} potential form, which can include the long-range charge relaxations with little or no extra cost beyond a static charge model \cite{ANiklasson21,ANiklasson21b}. 
The flexible charge models can be combined with linear atomic cluster expansions (ACE) \cite{Drautz19,Drautz20,Drautz21} for the short-range charge-independent parts of the interatomic potentials and for the atomic electronegativities. The linear ACE provides a tunable and systematically improvable accuracy in the description of local atomic properties. It also reduces the cost compared to alternative machine learning methods based on deep neural networks or kernel ridge regressions \cite{Drautz21,dusson_atomic_2022,ACE_Carbon_22}. The combined shadow molecular dynamics and ACE for flexible charge models therefore represent a natural choice for atomistic simulations that combines computational efficiency with accuracy. 

The Born-Oppenheimer potential of the polarizable flexible-charge models that we will use to describe the atomic interactions can be derived from a coarse-grained approximation of a range-separated first-principles density functional theory (DFT) \cite{hohen,KohnSham65,RParr89,RMDreizler90,JPerdew92,WKohn99}.
The range separation of the energy contributions between short and long-range interactions is important, because the ACE can only be used for atomic properties that are determined by their local environment, whereas non-local dependencies would lead to an explosion in the dimensionality, which in practice prohibits any meaningful parameterization.
Instead, the long-range potential and forces will be included from the known Coulomb interactions between  atom-centered flexible charges.

To derive the flexible charge models, the range-separated energy functional of DFT is expanded in fluctuations of the electron density around a reference density of overlapping electron densities of non-interacting neutral atoms.
This expansion can, at least in principle, be extended to any order.
Here we will use only the second-order expansion of the DFT energy functional. 
Our derivation of the Born-Oppenheimer potential for the coarse-grained flexible charge models is in many ways similar to the derivation of self-consistent charge density functional tight-binding (SCC-DFTB) theory
\cite{WHarrison80,MFoulkes89,DPorezag95,MElstner98,MFinnis98,TFrauenheim00,PKoskinen09,MGaus11,BAradi15,BHourahine20,ANiklasson21},
but instead of representing the electron density with single-particle orbitals, we will use a coarse-grained atom-centered multipole expansion for the charge fluctuations of each atom \cite{DMYork96,GTabacchi02,ANiklasson21}. The corresponding energy approximation defines the flexible charge model. The Born-Oppenheimer potential is given by the relaxed ground state for the flexible charges, which, in general, requires some iterative solver. 
This potential energy surface is then approximated by a \textit{shadow} Born-Oppenheimer potential that avoids the costly iterative charge relaxation process that normally is required to find the relaxed electronic ground state charges \cite{ANiklasson21}. 
The concept of shadow potentials (or shadow Hamiltonians) \cite{HYoshida90,CGrebogi90,SToxvaerd94,GJason00,ShadowHamiltonian,SToxvaerd12,KDHammonds20} is closely related to a backward error analysis and has been used frequently in the design of symplectic integration schemes that provide excellent accuracy and long-term stability.
Here the shadow potential concept for molecular dynamics simulations is applied to the polarizable charge models. The shadow Born-Oppenheimer potential is then included in the framework of XL-BOMD that enables highly efficient and stable simulations. Related methods have been developed previously \cite{KNomura15,AAlbaugh15,AAlbaugh17,AAlbaugh18}, but without the full benefit of a shadow potential and the latest generalized formulations of XL-BOMD that are applicable also to flexible charge models derived from coarse-grained orbital-free DFT \cite{ANiklasson21,ANiklasson21b}.

To demonstrate our shadow molecular dynamics and ACE scheme for flexible charge models we will use only the lowest-order monopole expansion of the local atomic charge fluctuations in its simplest form.
In this case the polarizable monopole model becomes a shadow molecular dynamics version of well-established charge equilibration (QEq) models \cite{WJMortier86,AKRappe91,ANiklasson21,Staacke_2022}.
To describe the short-range interaction potentials and the atomic electronegativities of this shadow XL-QEq model we use the linear ACE with SCC-DFTB theory as our reference.

To evaluate and demonstrate the applicability of ACE combined with our extended Lagrangian shadow molecular dynamics for the monopole charge equilibration (ACE+XL-QEq) model we will use 
two separate testbed systems. The first testbed is UO$_2$, which is a solid frequently used as nuclear fuel. The second testbed is a molecular system of liquid water (H$_2$O). 
These two testbed systems represent very different challenges for the ACE+XL-QEq scheme.

Our main goal is to present and demonstrate the general theoretical formulation of the shadow molecular dynamics scheme and ACE for flexible charge models and how the costly long-range charge relaxation problem can be avoided without any significant loss of accuracy. Providing ab-initio level interatomic potentials that can be used to study UO$_2$ or water under general conditions is not our goal.

Some of the key equations that summarize our theory are given in Eqs.\ (\ref{SE2_eta}, \ref{U2Min}, \ref{EOM_R}, \ref{EOM_n}, \ref{n_Int}). These equations describe the second-order shadow energy functional, the Born-Oppenheimer potential energy surface, the equations of motion, and the low-rank Krylov subspace approximation for the integration of the extended electronic degrees of freedom. 

The framework is implemented using LAMMPS \cite{SPlimpton95,thompson_lammps_2022} and a modified, developers version of the LATTE software packages \cite{LATTE_LAMMPS,LATTE_OLD,MCawkwell12,AKrishnapriyan17,Perriot2018-cg}. Atomic units are used throughout the text. ACE models of the energy and the electronegativity are enabled by the FitSNAP \cite{rohskopf2023fitsnap} and LAMMPS software packages. The training software, FitSNAP, may be used to train machine-learned descriptor-based interatomic potentials using a wide array of model forms and atomic environment descriptors. The FitSNAP training software leverages a computationally efficient LAMMPS back end for calculating descriptors and descriptor gradients used to train models. 
The ACE descriptors may be defined up to arbitrary body-order. This N-body truncation is used to systematically achieve an optimal cost to accuracy trade off for both short range energetics as well as models of electronegativity that vary with bond environment. Given their computational efficiency and versatility, ACE models are ideal candidates for coupled short-range, machine-learned potentials and long-range charge relaxation parameterizations. All of which may be trained with the open source version of FitSNAP and its corresponding libraries.

It is noted that ACEs may be constructed such that they depend on charge or charge transfer variables through the descriptors \cite{Drautz20}. The ability to generalize ACE models is powerful, but equilibrating charge during dynamic simulations of explicitly charge-dependent ACE models has not yet been addressed. To achieve this, we decouple the charge dependence from the ACE descriptors and instead model a charge dependent part of the systems with a flexible charge model derived from coarse-grained orbital-free DFT. The influence of bond environment on charge transfer is captured in a secondary ACE of the electronegativities in this flexible charge model, in addition to short-range energetics. As we will show, this additional flexibility over QEq models with fixed atom-dependent electronegativities can be important.

\section{Coarse-grained range-separated DFT}

Flexible charge models can be derived directly from first-principles Hohenberg-Kohn DFT \cite{hohen,RParr89,RMDreizler90}. Here we will use an approach similar to what originally was used to derive SCC-DFTB theory, but instead of using molecular orbitals to describe the electron density, we will use atom-centered multipole expansions. This gives us a \textit{coarse-grained} representation of DFT, without the effective single-particle details of an orbital-resolved Kohn-Sham DFT. To be able to separate long-range from short-range interactions we first present a simple range-separated form of Hohenberg-Kohn DFT.

\subsection{Range-separated Hohenberg-Kohn DFT}

In Hohenberg-Kohn DFT \cite{hohen,RParr89,RMDreizler90} the Born-Oppenheimer potential, $U_{\rm BO}({\bf R})$,
is given by a constrained minimization over all or physically relevant (or $v$-representable)
densities, $\rho({\bf r})$, where
\begin{equation}\label{BO1}
U_{\rm BO}({\bf R}) = \min_{\rho} \left\{ E({\bf R},\rho) ~\left \vert ~ \ \int \rho_{}({\bf r}) d{\bf r}  = N_e \right. \right\}.
\end{equation}
Here $E({\bf R},\rho)$ is a range-separated Hohenberg-Kohn energy functional of the electron density, which is defined by
\begin{equation}\label{RangeSep}
E({\bf R},\rho) = F_S[\rho] + E_{\rm H}[\rho] + \int V_{\rm ext}({\bf R,r})  \rho({\bf r})d{\bf r} + V_{nn}({\bf R}).
\end{equation}
Here $F_S[\rho] = F_{\rm HK}[\rho] - E_{\rm H}[\rho]$, which is the short-range (S) part of the universal Hohenberg-Kohn density functional, $F_{\rm HK}[\rho]$, after the long-range Hartree-term, 
\begin{align}
E_{\rm H}[{\bf \rho}] = \frac{1}{2} \iint \frac{\rho({\bf r}) \rho({\bf r'}) }{\vert {\bf r-r'}\vert} d{\bf r} d{\bf r'},
\end{align}
has been subtracted. The Hartree term is then added separately to $E({\bf R},\rho)$ in Eq.\ (\ref{RangeSep}) \footnote{Here we chose to ignore self-interaction corrections, which may affect the locality of the short-range functional $F_S[\rho]$.}.
$V_{\rm ext}({\bf R,r})$ is the external potential from the atomic ions at atomic positions, ${\bf R} = \{{\bf R}_i\}$, where ${\bf R}_I = [{R_I}_x, {R_I}_y, {R_I}_z]$, and $V_{nn}({\bf R})$ is the ion-ion repulsion.

\subsection{Charge-fluctuation expansion to second order}

To derive a coarse-grained expression for the range-separated energy functional, $E({\bf R},\rho)$, that is
useful to ACE and machine learning methods, we may follow the recipe of SCC-DFTB theory
\cite{MFoulkes89,DPorezag95,MElstner98,MFinnis98,DMYork96,GTabacchi02,ANiklasson21}, where the range-separated energy functional, $E({\bf R},\rho)$ is approximated by an expansion in the fluctuations, $\Delta \rho({\bf r}) = \rho({\bf r}) - \rho_{\rm ref}({\bf r})$, around some reference electron density, $\rho_{\rm ref}({\bf r})$. In general, we assume $\rho_{\rm ref}({\bf r})$ is given from the overlap of atom-centered electron densities of non-interacting neutral atoms. A second-order approximation, $E^{(2)}({\bf R},\Delta \rho)\approx  E({\bf R},\rho)$, is then given by 
\begin{widetext}
\begin{align}
E^{(2)}({\bf R},\rho) =& E({\bf R},\rho_{\rm ref}) + \int \left(\frac{\delta F_S[\rho]}{\delta \rho({\bf r})}\Big \vert_{\rho_{\rm ref}} + V_{\rm H}[\rho_{\rm ref}]({\bf r})+ V_{\rm ext}({\bf R,r})\right)\Delta \rho({\bf r})d{\bf r}  \nonumber \\ 
&~~~+ \frac{1}{2} \iint \Delta\rho({\bf r})\frac{\delta^2 F_S[\rho]}{\delta \rho({\bf r}) \delta \rho({\bf r'})}\Big \vert_{\rho_{\rm ref}}  \Delta \rho({\bf r'})  d{\bf r} d{\bf r'} + E_{\rm H}[\Delta \rho] \nonumber \\ 
=& V^{(0)}_S({\bf R}) + \int V_S^{(1)}({\bf r}) \Delta \rho({\bf r}) d{\bf r} + \frac{1}{2} \iint \Delta\rho({\bf r})V_S^{(2)}({\bf r,r'})  \Delta \rho({\bf r'})  d{\bf r} d{\bf r'}+ E_{\rm H}[\Delta \rho] \label{E2d},
\end{align}
\end{widetext}
which implicitly defines the interaction terms, $V^{(0)}_S({\bf R})$, $V_S^{(1)}({\bf r})$, and $V_S^{(2)}({\bf r,r'})$.
Additionally, $V_{\rm H}[\rho_{\rm ref}]({\bf r})$ is the Hartree potential, where
\begin{align}
V_{\rm H}[\rho_{\rm ref}]({\bf r}) =  \frac{\delta E_{\rm H}[\rho]}{\delta \rho({\bf r})} \Big \vert_{\rho_{\rm ref} }= \int \frac{\rho_{\rm ref}({\bf r'})}{\vert{\bf r-r'}\vert} d{\bf r'}.
\end{align}

The electron-density fluctuations can be expanded to any order, but here we will only use the
second-order approximation, $E^{(2)}({\bf R},\Delta \rho)$, as defined through the expression above. However, the theory is, at least in principle, straightforward to generalize also to higher orders.

The range-separated second-order energy functional, $E^{(2)}({\bf R},\Delta \rho)$, has several important
features. It clearly separates the terms dependent on the fluctuations in the electron density from the $\Delta \rho$-independent, $V_S^{(0)}({\bf R})$, term.
It is easy to see that $V_S^{(0)}({\bf R})$ is short-ranged if we assume that the reference density,
$\rho_{\rm ref}({\bf r})$, is given by overlapping short-range electron densities of non-interaction neutral atoms. $V_S^{(0)}({\bf R})$ should therefore
be well-suited to approximate with ACE and other machine learning methods, where atom-projected energy contributions are estimated from their local atomic environments. A sum of such atom-projected energies can then be used to represent $V_S^{(0)}({\bf R})$.  Also the second term, $V_S^{(1)}({\bf R,r})$, in Eq.\ (\ref{E2d}) is short-ranged.
This can be understood from the fact that the Hartree potential, $V_{\rm H}[\rho_{\rm ref}]({\bf r})$,
of overlapping neutral charge densities shields the
external potential from the ions, $V_{\rm ext}({\bf R,r})$, already at short distances,
and that $F_S[\rho]$ is short-ranged or ``nearsighted''. Also the third non-local term, $V_S^{(2)}({\bf r,r'})$, is short-ranged
because of the nearsightedness of $F_S[\rho]$.  The energy terms from $V_S^{(1)}({\bf R,r})$
and $V_S^{(2)}({\bf r,r'})$ should therefore, at least in principle, also be well-suited for ACE and machine learning methods that can capture the effects of the local environment of each atom.
All the remaining long-range interactions are captured separately by the electrostatic Hartree term, $E_{\rm H}[\Delta \rho]$, between the electron-density fluctuations, $\Delta \rho({\bf r})$, in the last term of Eq.\ (\ref{E2d}).
Addititional long-range dispersive interactions can be added separately, but will not be considered here.

\subsection{Coarse graining}

If we represent the charge fluctuations, $\Delta \rho({\bf r})$, in Eq.\ (\ref{E2d}), using molecular orbitals, we can recover the regular formulation of second-order SCC-DFTB theory, with the effective single-particle details of an orbital-resolved Kohn-Sham DFT.
Here we will instead approximate the fluctuations in the electron density, $\Delta \rho({\bf r})$, with more coarse-grained atom-centered local partial densities \cite{DMYork96,GTabacchi02,ANiklasson21} expressed in an atomic-orbital-like expansion,
\begin{align} \label {DeltaRho}
\Delta \rho({\bf r}) = \sum_{I,L \in \{nlm\}} \eta_{IL} \phi_{IL}({\bf r}),
\end{align}
where ${\boldsymbol \eta} = \{\eta_{IL}\}$ are the expansion coefficients.
The basis functions, $\{ \phi_{IL}({\bf r})\}$,
are local and centered around each atom $I$ at position, ${\bf R}_I$,
where $L \in \{nlm\}$ is a convenient multi-index for the radial and angular components. 
In this multipole expansion, the $l = 0$ and $m = 0$ basis functions are spherically symmetric (s-orbitals) and contain the net partial electron occupations surrounding each atom at ${\bf R}_I$. For all other orbitals $\int \phi_{IL}({\bf r}) d{\bf r} = 0, ~ l > 0$.
The choice of physically motivated basis functions can in principle allow for a natural $v$-representability of the density fluctuations.
However, there is no guarantee that the fluctuations avoid solutions where $\rho({\bf r}) < 0$ in $\Delta\rho({\bf r}) = \rho({\bf r}) - \rho_{\rm ref}({\bf r})$,
which would allow an unphysical amount of electron transfer between atoms, for example, where more than one electron is transferred from a single hydrogen atom.

Inserting Eq.\ (\ref{DeltaRho}) in the expression for $E^{(2)}({\bf R},\Delta \rho)$ in Eq.\ (\ref{E2d})
we get
\begin{align}\label{E2_eta}
&E^{(2)}({\bf R},{\boldsymbol \eta}) = V^{{(0)}}_S({\bf R}) + \sum_{IL} \chi_{IL} \eta_{IL} \nonumber \\
&+ \frac{1}{2} \sum_{IL,I'L'} \eta_{IL} \left(\Gamma_{IL,I'L'} + \gamma_{IL,I'L'}\right)\eta_{I'L'},
\end{align}
where we have used the integral expressions,
\begin{align}
\chi_{IL} =& \int V_S^{(1)}({\bf R,r}) \phi_{IL}({\bf r}) d{\bf r}, \label{ElNeg}\\
\Gamma_{IL,I'L'} =& \iint \phi_{IL}({\bf r}) V_S^{(2)}({\bf R,r}) \phi_{I'L'}({\bf r'}) d{\bf r}d{\bf r'},\\
\gamma_{IL,I'L'} =& \iint \frac{ \phi_{IL}({\bf r})  \phi_{I'L'}({\bf r'})}{\vert{\bf r-r'}\vert} d{\bf r}d{\bf r'}.
\end{align}
The first two set of terms, $\{\chi_{IL}\}$ and $\{\Gamma_{IL,I'L'}\}$, are both short-range, because $V_S^{(1)}({\bf R,r})$ and
$V_S^{(2)}({\bf R,r})$ are short range.
They are therefore, at least in principle, well-suited to approximations using ACE and machine learning methods. Although, they could also be calculated directly from DFT.

The Born-Oppenheimer potential, $U_{\rm BO}({\bf R})$, in Eq.\ (\ref{BO1}), can now be approximated by a coarse-grained second order Born-Oppenheimer potential, $U^{(2)}_{\rm BO}({\bf R}) \approx U_{\rm BO}({\bf R})$, which is given from the constrained minimization
of $E^{(2)}({\bf R},{\boldsymbol \eta})$ with respect to the expansion coefficients, $\{\eta_{IL}\}$, i.e.,
\begin{align}
U_{\rm BO}({\bf R}) 
& = \min_{{\boldsymbol \eta}} \left\{ E^{(2)}({\bf R},{\boldsymbol \eta}) ~\left \vert ~ \ \sum_{IL} \eta_{IL}\big \vert_{l = 0} =0   \right. \right\}. \label{UMIN}
\end{align}
The corresponding relaxed ground-state electron density distribution is given by
\begin{align}
&{\boldsymbol \eta}^{\rm min}  =  \arg \min_{{\boldsymbol \eta}} \left\{ E^{(2)}({\bf R},{\boldsymbol \eta}) ~\left \vert ~ \ \sum_{IL} \eta_{IL}\big\vert_{l = 0} = 0  \right. \right\}, \label{qmin}\\
&\Delta \rho_{\rm min}({\bf r}) =  \sum_{IL} \eta^{\rm min}_{IL} \phi_{IL}({\bf r}).
\end{align}
The constraint, $\sum_{IL} \eta_{IL}\vert_{l = 0}=0$, is included to make sure that no net charge is created or removed
in the fluctuations around the reference density, $\rho_{\rm ref}({\bf r})$, which here is assumed to
be the density of overlaping electron densities of non-interacting \textit{neutral} atoms. The representability condition of the total electron density is assumed to be automatically fulfilled by the choice of basis-set expansion functions, though, once again, there is no guarantee that we avoid unphysical amounts of charge transfer. 

The energy function, $E^{(2)}({\bf R},{\boldsymbol \eta})$, is quadratic in the density fluctuations and the relaxed ground state solution, ${\boldsymbol \eta}^{\rm min}$, is then given from
the solution of a linear system of equations,
\begin{align}
&\frac{\partial E^{(2)}({\bf R},{\boldsymbol \eta})}{\partial \eta_{IL}} = 0, \label{LSE}\\
& \sum_{IL} \eta_{IL}\big \vert_{l = 0} =0, \label{Q0}
\end{align}
where the charge constraint in Eq.\ (\ref{Q0}) can be included with a Lagrange
multiplier. For expansions of the energy functional using higher-order  expansions in the density fluctuations, the corresponding equations for the relaxed ground state electron distribution would become non-linear and have to be solved iteratively. The solution to Eqs.\ (\ref{LSE}) and (\ref{Q0}) gives us the optimized relaxed ground state, $\boldsymbol \eta^{\rm min}$, that defines the Born-Oppenheimer potential in Eq.\ (\ref{UMIN}). The molecular dynamics is then generated by Newton's equations of motion,
\begin{equation}
    M_I {\bf \ddot R}_I = - \nabla_I U_{\rm BO}({\bf R}),\label{NewtonEOM}
\end{equation}
where $\{M_I\}$ are the atomic masses for each atom $I$, and the dots denote the time derivatives.

Because of the long-range Coulomb interactions,
the system of equations, in Eq.\ (\ref{LSE}), is all-to-all. The long-range Coulomb interactions between flexible charges and their equilibration therefore increases the computational cost significantly compared to a charge-independent, short-ranged force field. For periodic systems, iterative solvers in combination with repeated Ewald summations (or the particle mesh Ewald method) can be used to solve Eqs.\ (\ref{LSE}) and (\ref{Q0}), but unless the solution is well converged, the calculated interatomic forces in Eq.\ (\ref{NewtonEOM}) may not be sufficiently accurate and conservative, which could invalidate a molecular dynamics simulation.  One of the key ideas behind the theory in this article is that we will use the concept of a 
\textit{shadow} dynamics based on a backward error 
analysis \cite{ANiklasson21}. As we will show in 
Sec.\ \ref{ShadowDynamics} below, this allows us to 
avoid the main part of the computational overhead associated with finding the relaxed ground state solution. This can be achieved without any significant loss of accuracy.

\section{Shadow energy functions and potentials} \label{ShadowDynamics}

\subsection{Shadow Molecular Dynamics}

The notion of a shadow dynamics provides a powerful concept
that helps us understand and design accurate and efficient integration schemes for molecular dynamics simulations
\cite{HYoshida90,CGrebogi90,SToxvaerd94,GJason00,ShadowHamiltonian,SToxvaerd12,KDHammonds20}.
A shadow dynamics is closely related to the technique of a backward error analysis. Instead of calculating approximate forces for an underlying exact potential energy surface (or Hamiltonian), we calculate exact forces, but for an underlying approximate ``shadow'' potential (or shadow Hamiltonian).
The \textit{simulated} dynamics will then be determined by the shadow potential. In this way important physical properties of the simulated dynamics such as time-reversiblity, total energy, and the phase-space area can be preserved, because the forces of the simulation are the exact conservative forces of the shadow potential energy surface, which closely follow the regular exact interatomic potential. In practice, such shadow molecular dynamics schemes become both more accurate and efficient compared to alternative regular techniques, in particular, the important long-term accuracy and stability is often superior.

The backward error analysis in terms of shadow Hamiltonians was originally applied to classical charge-independent molecular dynamics.
More recently, the concept was used to design a quantum-mechanical shadow Born-Oppenheimer molecular dynamics (QMD) that forms the theoretical underpinning of XL-BOMD \cite{ANiklasson07,ANiklasson08,MCawkwell12,JHutter12,LLin14,PSouvatzis14,ANiklasson17,ANiklasson21b}. 
Here we will use the same shadow molecular dynamics concept in the design of a shadow energy functional and shadow Born-Oppenheimer potential for the
coarse-grained second-order DFT model derived in the previous section. 
Thanks to the range separation of the different energy terms the theory can easily be combined with ACE and machine learning methods.
Our work is a natural extension, generalization and application of the theory and concepts that recently were introduced in Refs.\ \cite{ANiklasson21} and \cite{ANiklasson21b}.

\subsection{Shadow energy functions and potentials}

If we assume that we have some approximate ground state solution, ${\bf n} \approx {\boldsymbol \eta}^{\rm min}$,
we can linearize some of the terms in the energy function, $E^{(2)}({\bf R},{\boldsymbol \eta})$, in Eq.\ (\ref{E2_eta}), around ${\bf n} = \{n_{IL}\}$ without
too much loss of accuracy. In this way we can construct an approximate ${\bf n}$-dependent
shadow energy function, ${\cal E}^{(2)}({\bf R},{\boldsymbol \eta},{\bf n}) \approx E^{(2)}({\bf R},{\boldsymbol \eta})$, for example,
\begin{align} \label{SE2_eta}
&{\cal E}^{(2)}({\bf R},{\boldsymbol \eta},{\bf n}) =  V_S^{(0)}({\bf R}) + \sum_{IL} \chi_{IL} \eta_{IL}  \nonumber \\
&+ \frac{1}{2} \sum_{IL \ne I'L'} (2\eta_{IL} - n_{IL}) \Gamma_{IL,I'L'} n_{I'L'} \nonumber\\
&+ \frac{1}{2} \sum_{IL \ne I'L'} (2\eta_{IL}-n_{IL}) \gamma_{IL,I'L'} n_{I'L'} \nonumber \\
&+ \frac{1}{2} \sum_{IL = I'L'} \eta_{IL} \left(\Gamma_{IL,I'L'} + \gamma_{IL,I'L'}\right) \eta_{I'L'}.
\end{align}
Here we have linearized all terms except the diagonal terms in ${IL}$, which instead are kept to second order in $\eta_{IL}$. This is done to provide a unique solution to the constrained minimization problem that defines the shadow Born-Oppenheimer potential, 
\begin{align}
{\cal U}_{\rm BO}({\bf R,n}) = \min_{{\boldsymbol \eta} } \left\{ {\cal E}^{(2)}({\bf R},{\boldsymbol \eta},{\bf n}) ~\left \vert ~ \ \sum_{IL} \eta_{IL}\big \vert_{l = 0} =0 \right. \right\}.\label{U2Min}
\end{align}
With the minimization we here mean the lowest stationary solution.
The corresponding ${\bf n}$-dependent ground state solution for the charge fluctuations,
\begin{align}
{ \boldsymbol \eta}^{\rm min}[{\bf n}] =  \arg \min_{{\boldsymbol \eta} } \left\{ {\cal E}^{(2)}({\bf R},{\boldsymbol \eta},{\bf n}) ~\left \vert ~ \ \sum_{IL} \eta_{IL}\big \vert_{l = 0}  =0 \right. \right\}, \label{Eta_Min}
\end{align}
is then given from the solution of the system of linear equations,
\begin{align}
&\frac{\partial \left({\cal E}^{(2)}({\bf R,{\boldsymbol \eta},n})- \lambda \sum_{IL} \eta_{IL}\Big \vert_{l = 0}\right)}{\partial \eta_{IL} } = 0,\\
&  \frac{\partial \left({\cal E}^{(2)}({\bf R,{\boldsymbol \eta},n})- \lambda \sum_{IL} \eta_{IL}\big \vert_{l = 0}\right)}{\partial \lambda } = 0,
\end{align}
or
\begin{align}
&\chi_{IL} + \sum_{ I'L' \ne IL}  \left( \Gamma_{IL,I'L'} + \gamma_{IL,I'L'}\right)n_{I'L'} \nonumber \\
&~~~~~~~~+ \sum_{IL} \left(\Gamma_{IL,IL} + \gamma_{IL,IL}\right) \eta_{IL} - \lambda = 0,\\
& \sum_{IL} \eta_{IL}\Big \vert_{l = 0} = 0.
\end{align}
These equations have a simple analytical solution,
\begin{align} 
&\eta_{IL}^{\rm min}[{\bf n}] =  U_{IL}^{-1} \left(- \chi_{IL} - V^{\rm C}_{IL} + \lambda \right) \label{Ea}\\
& \lambda =  \frac{\sum_{IL} \left(\chi_{IL} + V^{\rm C}_{IL}\right) U^{-1}_{IL}}{\sum_{IL} U_{IL}^{-1}} \label{Eb},
\end{align}
which gives the relaxed ground-state solution of the charge fluctuations,
\begin{align} 
& \Delta \rho_{\rm min}[{\bf n}]({\bf r}) =  \sum_{IL} \eta^{\rm min}_{IL}[{\bf n}] \phi_{IL}({\bf r}).
\end{align}
Above we used the simplified notation,
\begin{align}
V^{\rm C}_{IL} = & \sum_{{I'L' \ne IL}} (\Gamma_{IL,I'L'}+ \gamma_{IL,I'L'})n_{I'L'}, \label{CoulombSum}\\
U_{IL} = &  \left(\Gamma_{IL,IL}+ \gamma_{IL,IL}\right).
\end{align}
No iterative solver with repeated sequential Coulomb or Ewald summations is needed. Only one single calculation of the Coulomb potential ${\bf V}^{\rm C} = \{V_{IL}^{\rm C}\}$ in Eq.\ (\ref{CoulombSum}) is needed. We have a simple
direct solution for the fully relaxed ground state  that is exact for the corresponding shadow energy function. 

Because $\eta_{IL}^{\rm min}[{\bf n}]$
is the exact solution (without any convergence problems), we have that
\begin{align}
\frac{\partial {\cal E}^{(2)}({\bf R},{\boldsymbol \eta},{\bf n})}{\partial \eta_{IL}} \Big \vert_{{\boldsymbol \eta}^{\rm min}} = 0.
\end{align}
This is important, because in the force evaluations, it means that any contributions from terms containing $\partial \eta_{IL} \big / \partial {{\bf R}_I}$ can be avoided. The shadow Born-Oppenheimer potential, ${\cal U}({\bf R,n})$, is only approximate, but in the context of a molecular dynamics simulation the calculated forces for the shadow potential are `exact' and easy to calculate. We therefore have no convergence problems, instabilities, or energy drift that can be caused by ill-converged, non-conservative forces.

The expression in Eq.\ (\ref{SE2_eta}) is only one particular choice to construct a shadow energy function. There are several alternative ways to construct a shadow energy function and corresponding Born-Oppenheimer potential, but they all have to fulfill certain requirements. There are three key conditions that, in general, need to be fulfilled. The first two conditions are
\begin{align}
&\left \vert E^{(m)}({\bf R},{\boldsymbol \eta})\ -{\cal E}^{(m)}({\bf R},{\boldsymbol \eta},{\bf n})\right  \vert \propto \left \vert {\boldsymbol \eta} - {\bf n} \right \vert^2,\\
&\left \vert \frac{\partial {\cal E}^{(m)}({\bf R},{\boldsymbol \eta},{\bf n})}{\partial n} \right \vert \propto \left \vert {\boldsymbol \eta} - {\bf n}\right \vert.
\end{align}
The third condition is that the equation
\begin{align}
& \frac{\partial  {\cal E}^{(m)}({\bf R},{\boldsymbol \eta},{\bf n})}{\partial {\boldsymbol \eta}} = 0,
\end{align}
(with additional net charge constraints) has a unique solution and can be solved in a simple direct way that avoids, e.g., a costly iterative procedure.
These three conditions are necessary but not sufficient and they therefore don't define a unique shadow energy function and potential.
However, the conditions guide and limit the possible design of the shadow
energy expressions and the corresponding shadow Born-Oppenheimer potentials. The shadow energy function in Eq.\ (\ref{SE2_eta}) is possibly the most simple and straightforward solution that fulfills the conditions above, but it may not be the most efficient choice. Alternative options will be considered elsewhere.

The error in the shadow Born-Oppenheimer potential is governed by the linearization of the energy function, $E^{(2)}({\bf R},{\boldsymbol \eta})$, around ${\bf n}$ such that
\begin{align}
    \left \vert {\cal U}_{\rm BO}({\bf R,n}) - U_{\rm BO}({\bf R}) \right \vert \propto \left \vert {\boldsymbol \eta}^{\rm min}[{\bf n}] - {\bf n}\right \vert^2.
\end{align}
To keep the error small we therefore need to keep ${\bf n}$ close to
${\boldsymbol \eta}^{\rm min}[{\bf n}]$, which keeps the residual
function, ${\bf f}({\bf n}) = {\boldsymbol \eta}^{\rm min}[{\bf n}] - {\bf n}$, small \footnote{This also means that ${\bf n}$ is close to the exact ground state of the regular Born-Oppenheimer potential, ${\boldsymbol \eta}^{\rm min}$.}. We can achieve this by propagating ${\bf n}$ as a dynamical field variable in addition to the atomic coordinates and velocities, where ${\bf n}(t)$ evolves through an extended harmonic oscillator that is centered around ${\boldsymbol \eta}^{\rm min}[{\bf n}]$ as the atoms are moving. This is one of the key ideas behind XL-BOMD.

\section{Extended Lagrangian Born-Oppenheimer Molecular Dynamics}

The idea of an extended Lagrangian framework goes back to Andersen's approach to molecular dynamics (MD) simulations at constant temperatures and pressures \cite{HCAndersen80,MParrinello80,SNose84}, where additional extended dynamical variables are introduced, besides the atomic positions and velocities, to enforce a given average temperature or pressure.
Car and Parrinello took the concept in a new direction and with a different purpose \cite{RCar85,DRemler90,GPastore91,FBornemann98,DMarx00,MTuckerman02,JHutter12}.
Instead of using the extended Lagrangian to introduce some external constraints for a classical
molecular dynamics simulation, they included the effective single-particle electronic wavefunctions as extended classical dynamical field variables in a first-principles molecular dynamics scheme, which originally was based on Kohn-Sham DFT. In this way it is possible to avoid the non-linear, quantum-mechanical, Kohn-Sham eigenvalue problem. Instead, the interatomic forces can be calculated on-the-fly from the constrained propagation
of the electronic degrees of freedom with its own mass and kinetic energy. Car-Parrinello molecular dynamics is a general framework that can be applied to a broad range of methods beyond the original Kohn-Sham plane-wave pseudopotential approach. It can be use, for example, in combination with orbital-free density functional theory \cite{GZerah92,JClerouin92,FLambert06},
polarizable force-field models \cite{MSprik88,MSprik90,DVanBelle92,GLamoureaux03}, methods using local atomic orbitals \cite{BHartke92} or density matrix formulations \cite{HBSchlegel01,SIyengar01,JMHerbert04},
and for correlated electron methods \cite{JLi16}.
Unfortunately, Car-Parrinello molecular dynamics has some practical shortcomings and often requires very short integration time steps.

XL-BOMD provides an efficient alternative to extended Lagrangian Car-Parrinello molecular dynamics and
as a general framework for molecular dynamics simulations it can also be adapted to different
levels of theory and descriptions of the electronic structure \cite{ANiklasson21b}.
It can be seen as a next generation extended Lagrangian first-principles molecular dynamics \cite{ANiklasson17}.
XL-BOMD is based on a different extended Lagrangian and leads to a different set of equations of motion compared to Car-Parrinello molecular dynamics.
The integation time step can be of the same order as in regular Born-Oppenheimer molecular dynamics, but XL-BOMD has an additional overhead in
the integration of the extended electronic degrees of freedom. Here we will use the XL-BOMD scheme together with our shadow Born-Oppenheimer potentials and ACE for flexible charge models.

\subsection{Extended Lagrangian}

The shadow Born-Oppenheimer potential, ${\cal U}^{(2)}_{\rm BO}({\bf R,n})$, in Eq.\ (\ref{U2Min}), has an error that is
of second order in the difference between the ground state solution ${\boldsymbol \eta}^{\rm min}[{\bf n}]$
and ${\bf n}$, i.e.\ in the residual function, ${\bf f}({\bf n}) = {\boldsymbol \eta}^{\rm min}[{\bf n}] - {\bf n}$. To keep the
error small during a molecular dynamics simulation, we need to update ${\bf n}$ as the atoms move. We can do this by propagating ${\bf n}$ as an extended dynamical vector variable, ${\bf n}(t)$, that is driven
by a harmonic oscillator that is centered around the ground state solution as the atoms are moving. This can be formulated
using an extended Lagrangian, which we define as
\begin{align} \nonumber
&{\cal L}({\bf R,{\dot R},n,{\dot n}})  =  \frac{1}{2} \sum_I M_I \vert {\bf R}_I\vert^2 - {\cal U}^{(2)}_{\rm BO}({\bf R,n}) \\
&+\frac{1}{2}\mu \sum_I {\dot \eta}_{IL}^2 - \frac{1}{2} \mu \omega^2 \sum_{IL,I'L'} \left(\eta^{\rm min}_{IL}[{\bf n}] - n_{IL}\right) \nonumber\\
&~~~~~~~\times T_{IL,I'L'}  \times \left(\eta^{\rm min}_{I'L'}[{\bf n}] - n_{I'L'}\right). \label{XL}
\end{align}
The extended harmonic oscillator in ${\cal L}({\bf R,{\dot R},n,{\dot n}})$ includes a fictitious
mass parameter, $\mu$, and oscillator frequency, $\omega$, which determines the time scale of the
extended dynamical variables ${\bf n}(t)$ and ${\bf \dot n}(t)$ for the electronic degrees of freedom. The atomic masses are given by $\{M_I\}$ for each atom $I$.
The harmonic oscillator includes a symmetric positive definite metric tensor, ${\bf T} = {\bf K}^{\rm T}{\bf K}$, where the kernel ${\bf K} = {\bf J}^{-1}$
is defined as the inverse of the Jacobian, ${\bf J}$, of the residual function,
\begin{align}
{\bf f}({\bf n}) = {\boldsymbol \eta}^{\rm min}[{\bf n}] - {\bf n},
\end{align}
i.e.,
\begin{align} 
&J_{IL,I'L'} =  \nonumber \frac{\partial \eta_{IL}^{\rm min}[{\bf n}] }{\partial n_{I'L'}} - \delta_{IL,I'L'}, \nonumber\\
 & =  -U_{IL}^{-1} \left( \Gamma_{IL,I'L'} + \gamma_{IL,I'L'} \right)_{IL \ne I'L'}\nonumber\\
  & + U_{IL}^{-1} \left( \frac{ \sum_{IL} U_{IL}^{-1} \left( \Gamma_{IL,I'L'} + \gamma_{IL,I'L'}  \right)}{\sum_{IL} U_{IL}^{-1} }\right)_{IL \ne I'L'} \nonumber \\
  &~~~~~~~- \delta_{IL,I'L'}. \label{Jacobian}
\end{align}
Here we used the exact analytical solution in Eqs.\ (\ref{Ea}) and (\ref{Eb}) to derived the full Jacobian expression. The kernel, ${\bf K} = {\bf J}^{-1}$, plays an important role in the time evolution of the extended electronic degrees of freedom, ${\bf n}(t)$. 
The kernel, ${\bf K}$, that defines the metric tensor, ${\bf T}$, is chosen such that the derived equations of motion become simple to evaluate and such that ${\bf n}(t)$ oscillates around a closer approximation to the exact regular ground state density. A more detailed explanation is given in Ref.\ \cite{ANiklasson21b}. It is easy to test on-the-fly during a simulation that the dynamics of ${\bf n}(t)$ closely follows the ground state density, which will be demonstrated in the examples.

\subsection{Equations of motion}

The equations of motion can be derived from Euler-Lagrange equations for the extended Lagrangian, ${\cal L}({\bf R, \dot R, n, \dot n})$ in Eq.\ (\ref{XL}). To stay consistent
with the Born-Oppenheimer approximation, which is based on the assumption of a time-separation
between the light and fast moving electrons and the much heavier and slowly moving nuclei, we derive the Euler-Lagrange equations with the same assumption of a time-scale separation, but now between the extended electronic
degrees of freedom, ${\bf n}(t)$ and ${\bf \dot n}(t)$, and the atomic coordinates and velocities. Once again, we assume that the nuclear motion is slow compared to the electronic degrees of freedom, ${\bf n}(t)$. This adiabatic separation \cite{ANiklasson21b} can be introduced by deriving the Euler-Lagrange's equations under the conditions that 
\begin{align}
&\lim \omega \rightarrow \infty,\\
&\lim \mu \rightarrow 0, \\
& \mu \omega ={\rm constant},
\end{align}
and asserting that $\vert \rho -n\vert \propto \omega^2$.
In this adiabatic mass-zero limit \cite{ANiklasson08,ANiklasson21,ANiklasson21b}
we get the coupled equations of motion,
\begin{align}
M_I{\bf \ddot R}_I  = & - \nabla_I {\cal U}_{\rm BO}({\bf R,n})\big \vert_{\bf n} \label{EOM_R}\\
{\bf \ddot n} = & - \omega^2 {\bf K}\left({\boldsymbol \eta}^{\rm min}[{\bf n}] - {\bf n} \right) \label{EOM_n}.
\end{align}

The first equation, Eq.\ (\ref{EOM_R}), is similar to regular Born-Oppenheimer molecular dynamics, Eq.\ (\ref{NewtonEOM}), but the shadow potential, ${\cal U}_{\rm BO}({\bf R,n})$, can be constructed in one shot with a simple analytical solution. No iterative solver or explicit matrix inversion is needed. In this way convergence errors are avoided and the exact conservative forces can be obtained for ${\cal U}_{\rm BO}({\bf R,n})$. It is important to note that this shadow Born-Oppenheimer potential, ${\cal U}^{(2)}_{\rm BO}({\bf R,n})$, is meaningful only in the context of XL-BOMD. In this case the interatomic forces in Eq.\ (\ref{EOM_R}) are calculated under constant ${\bf n}$, because ${\bf n}$ appear as dynamical variables. In a static non-XL-BOMD application ${\cal U}^{(2)}_{\rm BO}({\bf R,n})$ only corresponds to some generalization of a Harris-Foulkes-like energy expression in Kohn-Sham DFT \cite{JHarris85,MFoulkes89}, which only can be used to estimate ground-state energies, but not the interatomic forces \cite{ANiklasson14,ANiklasson17,Niklasson2023}. 

The second equation, Eq.\ (\ref{EOM_n}), is a harmonic oscillator equation for the extended degrees of freedom that describes the evolution of the approximate charge fluctuations, ${\bf n}(t)$, around which we performed the expansion for the shadow energy function, ${\cal E}^{(m)}({\bf R},{\boldsymbol \eta}, {\bf n})$. In the integration of this equation of motion we need to describe how the kernel, ${\bf K}$, acts on the residual. In section, \ref{KrylovLowRank}, we will show how this can be achieved by a preconditoned Krylov subspace approximation. In general, we do not need a very high accuracy in the approximation of the kernel to integrate the equations of motion in Eq.\ (\ref{EOM_n}). In this way the kernel appears more like a preconditioner. The kernel makes ${\bf n}(t)$ evolve around a closer approximation to the exact regular ground state, ${\boldsymbol \eta}^{\rm min}$, compared to the ground state, ${\boldsymbol \eta}^{\rm min}[{\bf n}]$, of the ${\bf n}$-dependent shadow potential.

In the derivation of the equations of motion above we assert an exact adiabatic separation in the limit of a vanishing mass and infinite frequency. 
The equations of motion are therefore exact in continuous time.
In practice, using finite integration time steps, the adiabatic separation between the electronic and nuclear degrees of freedom is only approximate. However, under normal conditions the adiabatic separation always seems to be sufficient \cite{ANiklasson21b}. Nevertheless, during a molecular dynamics simulations it is important to check the size of the residual function, ${\bf f}({\bf n}) = {\boldsymbol \eta}^{\rm min}[{\bf n}] - {\bf n}$, to make sure that the dynamical charge vector, ${\bf n}(t)$, always stays close to the ground state, ${\boldsymbol \eta}^{\rm min}[{\bf n}]$. We maintain a good adiabatic separation as long as the size of this residual remains small.

The adiabatic mass-zero limit above is related to the recent work by Bonella and co-workers \cite{ACoretti20,SBonella20,SBonella21}, where an adiabatic mass-zero constraint is enforced exactly by using a set of Lagrange multipliers for Car-Parrinello molecular dynamics. These Lagrange multipliers can then be determined iteratively in each time step. This mass-zero formalism is an efficient alternative to the original formulation of Car-Parrinello molecular dynamics and makes it possible to use integration time steps of the same order as in regular Born-Oppenheimer molecular dynamics.

\subsection{Integrating the equations of motion using a preconditioned Krylov subspace approximation} \label{KrylovLowRank}

In the integration of the equations of motion for the nuclear degrees of freedom in Eq.\ (\ref{EOM_R}) we can use the standard leapfrog velocity Verlet scheme.
The integration of the harmonic oscillator equation of motion in Eq.\ (\ref{EOM_n}) for the
extended electronic degrees of freedom, on the other hand, requires some care. 
Typically, the Verlet integration of the extended electronic equations of motion needs to be modified to include some weak form of dissipation to keep ${\bf n}(t)$ synchronized with the trajectories of the atomic positions
\cite{ANiklasson09,PSteneteg10,GZheng11,AOdell09,AOdell11,VVitale17,ANiklasson21b}. Other alternative integration schemes proposed by Head-Gordon and co-workers could also be used \cite{ILeven19}.

In addition to the modified Verlet integration, we also need to approximate
the kernel, ${\bf K}$, and how it acts on the residual function,
${\bf f(n)} = {\boldsymbol \eta}^{\rm min}[{\bf n}] - {\bf n}$, in the electronic equations of motion, Eq.\ (\ref{EOM_n}). This can be achieved
with a preconditioned Krylov subspace approximation \cite{ANiklasson20}, where the kernel acting on the preconditioned residual is given by a low-rank approximation. In this case we can express Eq.\ (\ref{EOM_n}) as 
\begin{align} \label{n_Int}
{\bf \ddot n} = - \omega^2 \left(\sum_{kl} {\bf v}_k {\widetilde M_{kl}} {\widetilde {\bf f}_{{\bf v}_l}} \right){\bf K}_0\left({\boldsymbol \eta}^{\rm min}[{\bf n}] - {\bf n} \right).
\end{align}
The different terms, $\{{\widetilde M_{kl}}\}$, $\{{\bf v}_k\}$, $\{{\widetilde {\bf f}_{{\bf v}_l}}\}$ of the low-rank kernel approximation and the preconditioner, ${\bf K}_0 \approx {\bf J}^{-1}$, are defined in the appendix. This technique provides an efficient method to approximate the kernel in the integration of the electronic degrees of freedom.

\section{Monopole Charge Equilibration model}

The theory presented above is fairly general and is applicable to a wide range of polarizable models based on a second-order expansion in the charge fluctuations of the range-separated Hohenberg-Kohn energy functional, where the local energy terms can be represented using ACE and machine learning methods. To demonstrate our theory we will use the same second-order expansion of the Hohenberg-Kohn energy functional. However, we will use only the lowest monopole order, where $l = 0$, with only one fixed radial function ($n=1$) for the coarse-grained atom-centered expansion of the density fluctuations, $\Delta \rho({\bf r})$ in Eq.\ (\ref{DeltaRho}), for the shadow energy function, ${\cal E}^{(2)}({\bf R},{\boldsymbol \eta},{\bf n})$, in Eq.\ (\ref{SE2_eta}). To simplify the notation we then drop all the $L$ indices.
We will further assume that the potential parameters $\Gamma_{I,I'}$ are local, i.e.\ $\Gamma_{I,I'} = \Gamma_{I,I'} \delta_{I,I'}$. In this case the shadow energy function, ${\cal E}^{(2)}({\bf R},{\boldsymbol \eta},{\bf n})$ in Eq.\ (\ref{SE2_eta}), becomes
\begin{align}
&{\cal E}^{(2)}({\bf R},{\boldsymbol \eta},{\bf n}) =  V_S({\bf R}) + \sum_{I} \chi_{I} \eta_{I} \nonumber \\
&+ \frac{1}{2} \sum_{I'}^{I' \ne I} (2\eta_{I}-n_{I}) \gamma_{I,I'} n_{I'}
+ \frac{1}{2} \sum_{I}  U_I \eta_{I}^2, \label{E2_QEQ}
\end{align}
where we use the combined Hubbard-U like parameter, $U_I = \Gamma_{I,I} + \gamma_{I,I}$, for the last on-site energy term.
This simplified polarizable energy function for $L=0$ is a shadow version of well-established energy functions in second-order flexible QEq models \cite{WJMortier86,AKRappe91,DMYork96,ANiklasson21,Staacke_2022}.
The Coulomb potential is given by $V_I^C = \sum_{I'}^{I'\ne I} \gamma_{I,I'} n_{I'}$. Notice that the expansion coefficients, ${\boldsymbol \eta} = \{\eta_I\}$, describe the occupation or the population of the partial atomic net charges rather than the charge itself. The $\{\chi_I\}$ terms that determine the linear dependency of the energy with respect to the atomic partial charges therefore correspond to the atomic electronegativities, but with the opposite sign. We will not  make those sign distinctions for  ${\boldsymbol \eta}$ and $\{\chi_I\}$ in our discussion. We will simply refer to them as partial atomic charges and electronegativities.

The different terms, $V_S({\bf R})$ and $\{\chi_I\}$, of the QEq model in Eq.\ (\ref{E2_QEQ}) will be fitted to reference data using the linear ACE. In principle, they could also be calculated directly from DFT and other machine learning methods could be used. However, the computational cost, in general, would be much higher.
We will use SCC-DFTB theory as our reference ground truth. SCC-DFTB theory, as implemented in the LATTE electronic structure package, is parameterized from DFT \cite{LATTE_OLD,MCawkwell12,AKrishnapriyan17}. The SCC-DFTB reference data that we will use to parameterize the QEq model can therefore be seen as an intermediate step between first-principles electronic structure theory and the QEq model. Because of this indirect parameterization from first-principles data we may lose some accuracy and fidelity. However, our goal is not to provide any ab-initio level interatomic potentials for accurate molecular dynamics simulations under general conditions.
The goal is instead to qualitatively demonstrate our shadow molecular dynamics and ACE methodology for a simple second-order flexible charge model and show how the costly long-range charge relaxation problem can be avoided without any significant loss of accuracy. 

The first two terms, $V_S({\bf R})$ and $\chi_{I} \equiv \chi_{I}({\bf R})$, in Eq.\ (\ref{E2_QEQ}) are short-range and therefore well suited for a parameterization based on their local atomic environments using the ACE. For the third and fourth term we use a screened Coulomb interaction of overlapping charge densities, with the on-site terms equal to predetermined Hubbard-U (or chemical hardness) parameters, $U_I$, for each atom type and where the long-range behavior is given by the bare Coulomb interaction decaying as $1/\vert{\bf R-R'}\vert$. To account for periodic boundary conditions we use the Ewald summation method \cite{MAllen90}. 

The approximate, $n=1$ and $l=0$, shadow energy function, ${\cal E}^{(2)}({\bf R},{\boldsymbol \eta},{\bf n})$, in Eq.\ (\ref{E2_QEQ}) defines the shadow Born-Oppenheimer potential, ${\cal U}^{(2)}({\bf R,n})$, from the constrained minimization in Eq.\ (\ref{U2Min}). This potential energy surface is then used for the integration of the equations of motion, Eqs.\ (\ref{EOM_R}) and (\ref{EOM_n}), that generates the molecular trajectories of the XL-BOMD simulation. In the examples below we will use a preconditioner ${\bf K}_0$ that is calculated from ${\bf J}^{-1}$, Eq.\ (\ref{Jacobian}), in the first initial time step. The preconditionder can then be updated in some chosen time interval, e.g.\ every 1,000 or 10,000 time step, or it could be kept constant throughout the simulation. This choice of preconditioner is then combined with the low-rank Krylov subspace approximation in Eq.\ (\ref{n_Int}) for the integration of the equations of motion for ${\bf n}(t)$ (see Appendix).

\section{Atomic cluster expansion for UO$_2$ and water}

To parameterize the shadow energy function in Eq.\ (\ref{E2_QEQ}) we use the linear ACE \cite{Drautz19,Drautz20,Drautz21}. There are many alternative methods based on various machine learning techniques \cite{JBehler07,Gabor10,MRupp12,Anatole15,SGoedecker15,APThompson15,Faber17,KSchutt17,JHan18,JSSmith19,NLubbers18,TMiller20,Drautz21,Parsaeifard_2021,MLReview22}, but the ACE allows for a systematic improvement to any chosen order in the many-body interactions and in the radial and angular resolution of each atomic environment. 
The linear ACE is also a computationally efficient method \cite{ACE_Carbon_22} that can be seen as a generalization of several alternative machine learning descriptors \cite{Drautz21}. 

The shadow Born-Oppenheimer potential, Eq.\ (\ref{U2Min}), is given by the constrained minimization of the shadow energy function of the QEq model in Eq.\ (\ref{E2_QEQ}), where the short-ranged potential, $V_S({\bf R})$, and the electronegativities, $\chi_{I}(R)$, are parameterized with ACE. This shadow Born-Oppenheimer potential is then combined with XL-BOMD, with the dynamics given by the equations of motion in Eqs.\ (\ref{EOM_R}) and (\ref{EOM_n}). To demonstrate this ACE+XL-QEq shadow molecular dynamics scheme we will use two target testbed systems with periodic boundary conditions: crystalline UO$_2$; and liquid water, H$_2$O, as described by SCC-DFTB theory. We will demonstrate how the ACE+XL-QEq scheme provides a computationally efficient approach for molecular dynamics simulations. The ACE+XL-QEq scheme avoids the cost of an iterative optimization process, which is normally required to find the relaxed and equilibrated ground states charges prior to each force evaluation. This increased efficiency comes without any significant loss of accuracy. We will also show the significance of having a flexible parameterization of the atomic electronegativities. Without this flexibility, which is provided by our linear ACE, it is often impossible to accurately capture the correct charge distribution in many important scenarios. QEq models with fixed electronegativities for each atom type are therefore often inadequate.

We first present some background on the linear ACE model used in the parameterization of the shadow energy function in Eq.\ (\ref{E2_QEQ}). We then discuss some of the details in the ACE optimizations for solid UO$_2$, and water, which then are used in molecular dynamics simulations to demonstrate the ACE+XL-QEq shadow molecular dynamics scheme. 

\subsection{Atomic cluster expansion}

We use the ACE parameterization to model the charge-independent, short-ranges part of the force field, $V_S({\bf R})$, and the electronegativities, $\chi_{R}$, in Eq.\ (\ref{E2_QEQ}). These models are expansions of the per-atom properties in terms of ACE descriptors, $\{B^I_{\boldsymbol{\mu n l }}\}$, for each atom $I$ at position ${\bf R}_I$. The descriptors are rotationally and permutationally invariant by construction, and are indexed on radial function indices, $n_I$, angular component index, $l_I$, and
a chemical index, $\mu_I$, as described by Drautz in Ref.\ \cite{Drautz19}. A permutation-adapted approach is used to remove redundant descriptors above 4-body terms \cite{dusson_atomic_2022,goff_permutation-adapted_2022}. An atomic and site-dependent property such as the atomic electronegativity, $\chi_I \equiv \chi_I({\bf R})$, can then be described with a linear expansion of the descriptors, $B^I_{\boldsymbol{\mu n l }}$, up to arbitrary accuracy as,
\begin{equation}
    \chi_I({\bf R}) = \sum_{\boldsymbol{\mu n l }} c_{\boldsymbol{\mu n l }} B^I_{\boldsymbol{\mu n l }}({\bf R}),
    \label{eq:linear_expansion}
\end{equation}
where $c_{\boldsymbol{\mu n l }}$ are expansion coefficients for each atom type to be trained on  either the ground truth values of the
per-atom electronegativities, $\chi_I({\bf R})$, or on the ground truth values of the partial atomic charges generated by the electronegativities \cite{TWKo20}. While the first approach works well for UO$_2$, the indirect training on the partial charges works better for water.
The charge-independent potential energy term, $V_S({\bf R})$, is represented as a sum over local energy terms, $E_{\rm S}^I({\bf R})$, for each atomic site, where
\begin{align}
    E_{S}^I({\bf R}) =& \sum_{\boldsymbol{\mu n l }} c_{\boldsymbol{\mu n l }} B^I_{\boldsymbol{\mu n l }}({\bf R}),
    \label{eq:charge_ind_en}\\
    V_S({\bf R}) =& \sum_I E_S^I({\bf R}) \label{eq:charge_ind_pot}.
\end{align}

The training data set for both UO$_2$ and H$_2$O are built from properties that are generated from electronic structure calculations using the LATTE software package \cite{LATTE_OLD,MCawkwell12,AKrishnapriyan17}
based on SCC-DFTB theory \cite{MFoulkes89,DPorezag95,MElstner98,MFinnis98,TFrauenheim00,BHourahine20}. In the case of UO$_2$, the training data was generated from a finite temperature QMD simulation using a 96-atom UO$_2$ supercell, as well as a 95-atom UO$_2$ supercell with a Uranium vacancy. From these simulations, 200 and 50 frames respectively, were extracted from the QMD trajectory. The computed energy, forces, partial atomic charges, and electronegativities were extracted from each frame and used to build the training set. The atomic electronegativities were determined from the calculated partial charge distribution, where the Hubbard-U parameters, $\{U_I\}$, are kept from the SCC-DFTB reference. The electronegativities are given by $\chi_I = -V_{I}^{C} - U_I\cdot\eta^{\rm{min}}_{I}[\mathbf{n}] + \mu^{\rm ref.}$. 
Training the ACE parameters on the calculated ground truth electronegativities must include an additional unknown reference chemical potential parameter, $\mu^{\rm ref.}$, because the electronegativies are only determined up to an unknown constant shift for each new atomic configuration. Alternatively, we may optimize the ACE parameterization directly to the calculated partial atomic charges, where the condition of charge neutrality automatically determines the chemical potential for each training structure. 
While both approaches provide electronegativity models that produce stable MD trajectories, the latter seems to provide a more accurate parameterization using the same amount of training data, especially for the water system, where the values of the chemical potential, $\mu^{\rm ref.}$, can have a large variation. 

The charge-independent energy, $V_S({\bf R})$, was trained using per-atom average energies and per-atom forces using Bayesian Automatic Relevance Determination Regression (ARDR). The ARDR method is used to find an optimally sparse set of expansion coefficients, while minimizing overfitting \cite{wipf_new_2007}. The starting set of features before the Bayesian ARDR regression includes 320 ACE descriptors, for body orders between two to five. The maximum radial and angular quantum numbers used for the ACE descriptors for UO$_2$ are $n_{max}=16$ and $l_{max}=2$, respectively. For UO$_2$, these methods were sufficient to produce a charge-independent potential, $V_S(\mathbf{R})$, with energy and force root mean square error (RMSE) of 1.6 meV/atom and 148 meV/($\AA \cdot atom$), respectively. A Ziegler-Biersack-Littmark (ZBL) potential with an inner and outer cutoff of 0.6 and 2.5 $\AA$ for all bond types, in conjunction with hard-core repulsions below 0.1 $\AA$ is added as a reference potential to capture short-range repulsions \cite{ziegler1985stopping}. 

For water, using body orders between two to five, the starting set of features before the Bayesian ARDR regression includes 260 ACE descriptors. The maximum radial and angular quantum numbers used for the ACE descriptors for H$_2$O are $n_{max}=22$ and $l_{max}=2$, respectively. The charge-independent potential, $V_S({\bf R})$, for water was trained to achieve energy and force RMSE of 2.9 meV/atom and 122 meV/($\AA \cdot atom$), respectively. A Ziegler-Biersack-Littmark (ZBL) potential, with an inner and outer cutoff of 0.1 and 0.45 $\AA$ for all bond types, in conjunction with hard-core repulsions below 0.1 $\AA$ is added as a reference potential to capture short-range repulsions in water as well.

With the ultimate goal of running molecular dynamics simulations while including variable electronegativity per atomic site, the electronegativity models had to be optimized such that they produced stable dynamics when combined with the previously developed short-range potentials. For this reason, the electronegativity models were allowed to have different hyperparameters than those used for the parameterization of $V_{\rm S}({\bf R})$. The electronegativity models for UO$_2$ and water were trained on the same training data as for the charge-independent energy and forces, but using the electrostatic energies and forces. For UO$_2$, we also used two to five body ACE descriptors, but with the maximum radial and angular quantum numbers of $n_{max}=8$ and $l_{max}=2$, respectively. Both for the short-range potential and the electronegativity model, the maximum radial cutoff of the ACE descriptors was set to 5.5 $\AA$ for UO$_2$. The optimized electronegativity model predicts electronegativities of UO$_2$ within an RMSE of 106 mV. This should be compared to the difference in electronegativities between U and O, which is 4.45 V as given by our DFTB reference data. The combined short-range potential and site-dependent electronegativity-driven ACE+XL-QEq scheme result in a stable dynamics for UO$_2$ not only in pristine crystals and under ambient conditions, but also for systems with point defects and at high-temperatures. 

For water we used two different eletronegativity models, in the first model we trained directly on the electronegativities generated from the training data, and in the second model by optimizing the ACE coefficients to generate the correct partial atomic net charges \cite{TWKo20}.
The first electronegativity model for water was trained in a way similar to that of UO$_2$, using electronegativities calculated by, $\chi_I = -V_{I}^{C} - U_I\cdot\eta^{\rm{min}}_{I}[\mathbf{n}] + \mu^{\rm{ref}}$. This electronegativity model for water was trained with 210 SCC-DFTB frames from a 300-atom molecular dynamics simulation. The resulting RMSE in the electronegativity model was 0.221 V for this training set. Although this is relatively high compared to the accuracy for the UO$_2$ electronegativity model, it can accurately predict the electronegativities of bulk water on average. 
The second model trained on the partial atomic net charges gave a more accurate ACE parameterization. The RMSE for the second electronegativity model was only 0.010 V. This second water model is more accurate compared to the first model, which showed a tendency of enhanced autoionization under ambient conditions.  The ACE descriptors in the water models have maximum radial cutoffs of 3.652 $\AA$. A cutoff of 3.652 $\AA$ was found to be optimal within a test range from 3.0 to 6.0 $\AA$ for linear electronegativity ACE models.

Our ACE parameterizations of UO$_2$ and water can be improved using a larger and more diverse set of training data, especially for water. We could also replace the SCC-DFTB reference data with properties calculated directly from high-level \textit{ab initio} theory, e.g. including important electron correlation effects beyond regular DFT for the case of UO$_2$ fuels \cite{VIAnisimov91,VIAnisimov97,AILichtenstein95,SLDudarev98,HIdriss10,HHeming13,HKulik15}.
The main purpose of our demonstration, however, is not to provide new and accurate interatomic potentials for UO$_2$ fuel or water. Our main goal is only to show how the ACE+XL-QEq shadow molecular dynamics scheme generates stable trajectories that reproduce the charges and potential energy surface of the corresponding exact Born-Oppenheimer model. In this respect our two testbed systems only represent two archetypal model problems -- one for a solid and the other for a liquid.

\subsection{LAMMPS implementation}

The models are trained using the LAMMPS molecular dynamics code and the FitSNAP software package \cite{SPlimpton95,zuo_performance_2020,thompson_lammps_2022}. The FitSNAP software package may be used to train machine-learned interatomic potentials using LAMMPS as a backend to compute descriptor values for each atomic configuration. Bayesian-learning methods used for training the ACE models are those implemented in the SKLearn library \cite{scikit-learn}.

The ACE+XL-QEq molecular dynamics simulations were carried out with LAMMPS and a developers version of the LATTE electronic structure code along with the computationally efficient ACE kernels as implemented by Lysogorskiy et al. \cite{Drautz21}.
The flexible ACE electronegativities were evaluated with the newly implemented ``compute pace" (LAMMPS compute for ACE descriptors) \cite{goff_permutation-adapted_2022}. This LAMMPS compute relies on the same kernels as previous implementations of ACE, and it shares the same computational efficiency \cite{Drautz21}. The interatomic forces in these simulations also accounted for the force contributions from the gradients of the site-dependent electronegativities with respect to their atomic positions.

\begin{figure}
    \centering
    \includegraphics[width=0.50\textwidth]{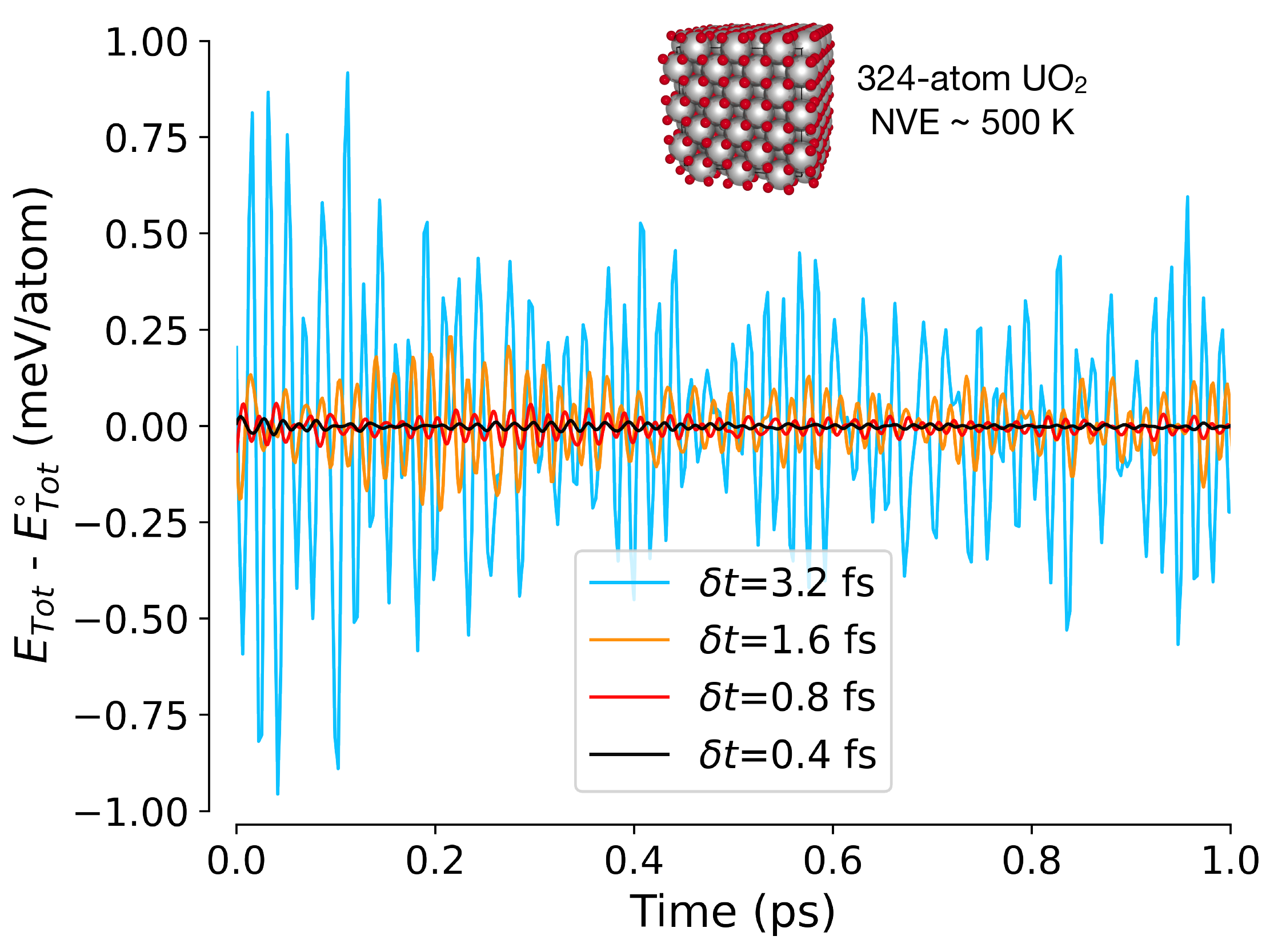}
    \caption{Microcanonical (NVE) shadow molecular dynamics simulations using the ACE+XL-QEq scheme of a 324-atoms supercell of UO$_2$. The fluctuations in the total energy (deviation from mean) for different values of time integration step highlight the approximate $\delta t^2$ scaling with the amplitude of the oscillations in the total energy and the stability. The statistical temperature fluctuates around about 500 K.}
    \label{fig:dt_scaling}
\end{figure}

\subsection{ACE+XL-QEq shadow molecular dynamics}

To demonstrate the ACE+XL-QEq shadow molecular dynamics scheme, we will use two test systems: UO$_2$ and water. These test systems are chosen to illustrate the broad range of applicability of the ACE+XL-QEq shadow molecular dynamics approach.

\begin{figure*}{h}
    \centering
    \includegraphics[width=0.70\textwidth]{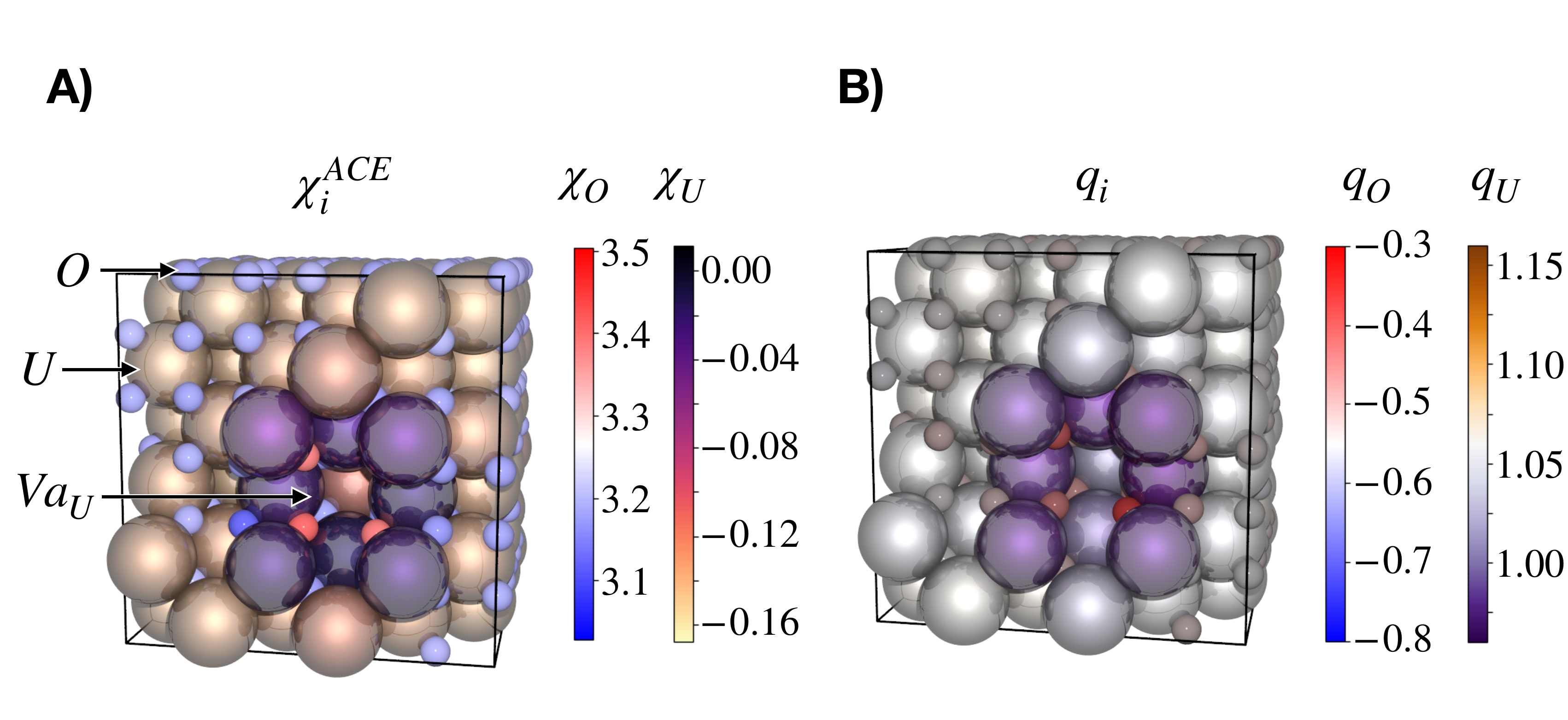}
    \caption{Left panel (A): molecular dynamics snapshot of the flexible ACE electronegativities of a 323 atom UO$_2$ supercell containing a single U vacancy, $Va_U$. Right panel (B): Same molecular dynamics snapshot of the partial charge distributions near the uranium vacancy.}
    \label{fig:eneg_to_charge}
\end{figure*}

\subsubsection{UO$_2$}

To demonstrate the stability of the combined short-range potential and electronegativity model for UO$_2$ using the ACE+XL-QEq shadow molecular dynamics scheme, we perform microcanonical NVE simulations of a 324-atom super cell of UO$_2$ with periodic boundary conditions  shown in Fig. \ref{fig:dt_scaling}. The system is initialized with out-of-equilibrium configurations and given a range of initial velocities corresponding to ionic statistical temperatures between 1000 K to 4000 K, resulting in average simulation temperatures in the range of 500 K to 2000 K. 
For all the  UO$_2$ molecular dynamics simulations, the preconditioner is updated every 500 time steps, though we have observed that the simulations are not very sensitive to the frequency of preconditioner updates.  The kernel acting on the preconditioned residual in the integration of the electronic degrees of freedom, Eq.\ (\ref{n_Int}), is updated with a rank-2 update on average during the simulation. The relaxed ground state charge distribution that determines the shadow Born-Oppenheimer potential is given by the exact analytical solution in Eq.\ (\ref{Ea}), which only requires one single construction of the Coulomb potential, Eq.\ (\ref{CoulombSum}). 
As expected for the Verlet integration scheme used here, the fluctuations of the total energy, $E_{\rm tot.} =  E_{\rm pot.} + E_{\rm kin.}$, scale with the square of the size of the integration time step, $\delta t$.

The flexible electronegativities parameterized by the linear ACE result in an accurate prediction of the Coulomb energy.  For the 96-atom UO$_2$ structure evolved in an NVE reference simulation using SCC-DFTB, the average Coulomb energy is -152 meV/atom.  In comparison, when the same 96-atom structure is evolved with the ACE+XL-QEq shadow molecular dynamics simulation, the average Coulomb energy is -151 meV/atom, in close agreement with the reference data.
The shadow molecular dynamics simulations obtained with the ACE+XL-QEq model are stable, with minimal systematic drift in the total energy, and allow for simulations at a significantly reduced computational cost relative to a regular QEq scheme.  A regular QEq scheme, in general, would require costly and tightly converged iterative solutions for the ground state charges prior to each force evaluation. 

By experimenting with artificially introduced point defects in the UO$_2$ system, we were able to further assess the ACE+XL-QEq model.  After introducing a uranium cation vacancy (artificially removing a U atom), the charge redistribution around the vacancy is driven by a significant change in the values of the flexible electronegativities near the vacancy center. 
\begin{figure}
    \centering
    \includegraphics[width=0.50\textwidth]{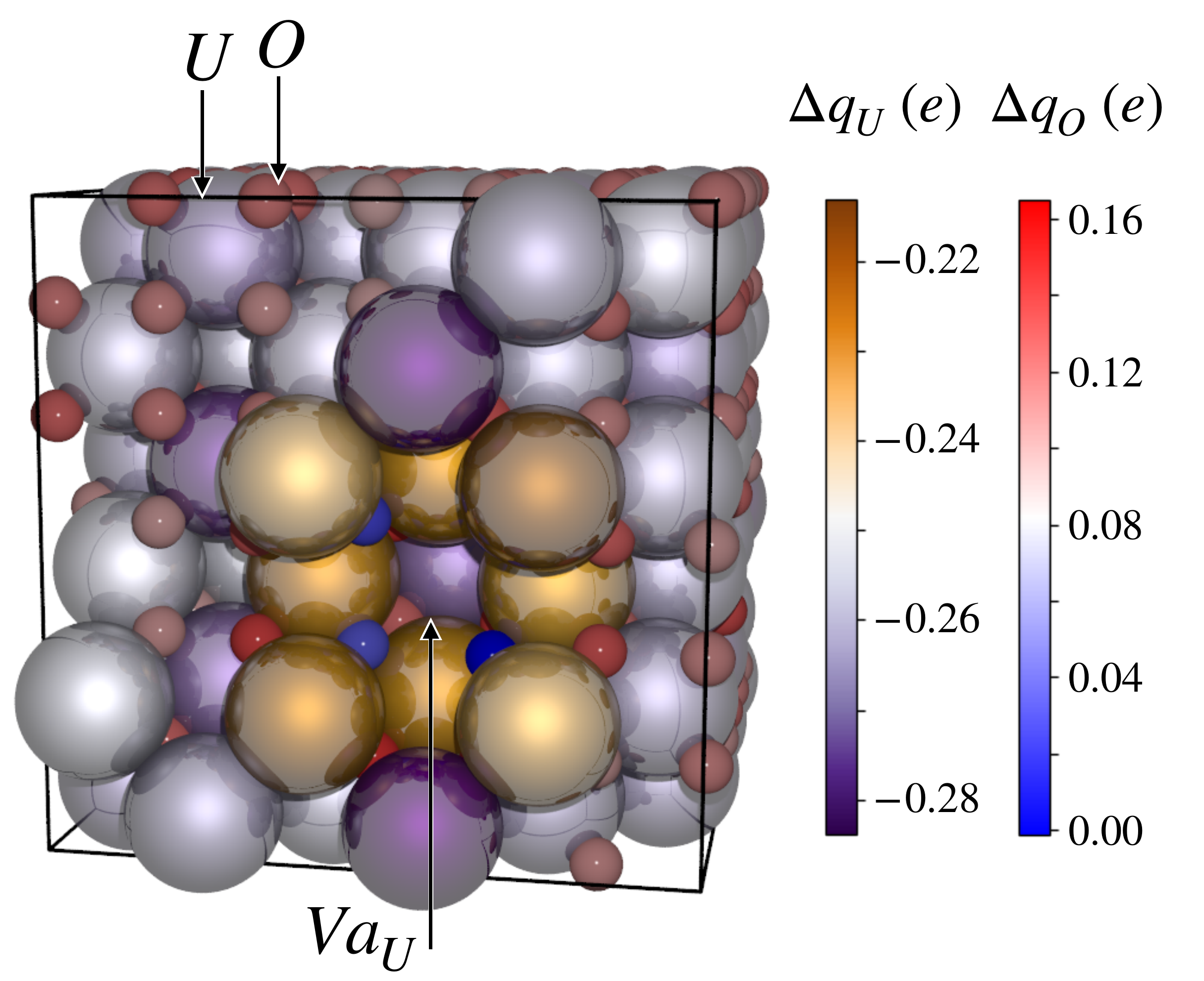}
    \caption{Time-averaged differences for U and O, $\Delta q_U$ and $\Delta q_O$, during a molecular dynamics simulation between the partial charges obtained with ACE parameterized flexible electronegativities and those obtained using fixed electronegativities for the UO$_2$ system with a U vacancy defect.}
    \label{fig:compare_uo2_vac}
\end{figure}
\begin{figure*}
    \centering
    \includegraphics[width=0.95\textwidth]{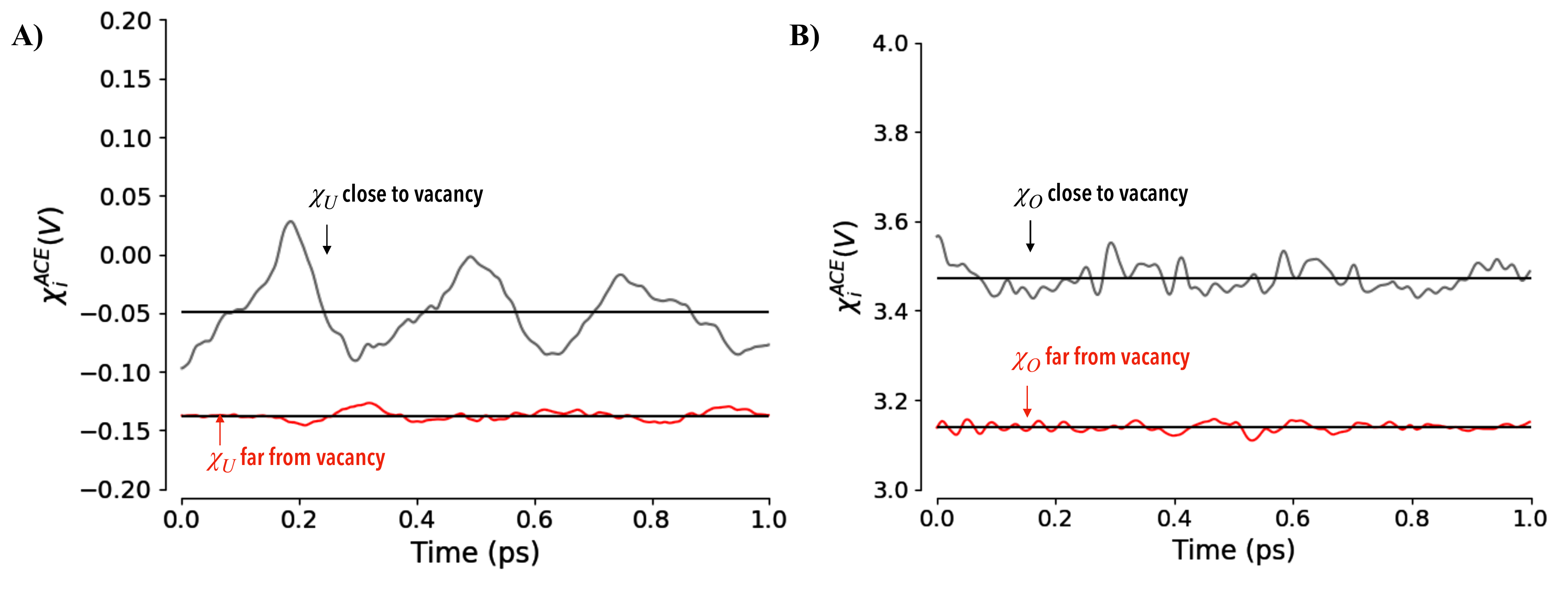}
    \caption{The electronegativities during an NVE simulation of the 323 atom UO$_2$ supercell with a vacancy, in Fig.\ \ref{fig:eneg_to_charge}, for uranium (A) and oxygen (B) atoms, both near and far from the vacancy center, ${Va}_U$. Horizontal lines represent time-averaged values of the electronegativity over the simulation window pictured. }
    \label{fig:eneg_plot}
\end{figure*}
This is displayed in Fig. \ref{fig:eneg_to_charge}, where the flexible atomic electronegativities are shown on the left side (A) for a frame (snapshot) of a molecular dynamics simulation, along with the charge distribution, on the right side (B), for the same frame. 
The vacancy creates a perturbation in the UO$_2$ supercell which is screened by a charge redistribution in the closest atomic shells surrounding the vacancy, as seen in Fig. \ref{fig:eneg_to_charge} (B).  If we used fixed electronegativities for each atom, we would not be able to fully capture these changes.  This is demonstrated in Fig.\ \ref{fig:compare_uo2_vac}, which shows the difference in partial charges resulting from using a flexible ACE electronegativity model with those from a fixed electronegativity model. Figure \ref{fig:compare_uo2_vac} also reveals that the flexible ACE model increases the positive charge on the third-nearest neighbors (oxygen) of the uranium vacancy and slightly increases the charge of the next-nearest neighbors (uranium).  This is in agreement with SCC-DFTB reference data for the charge distributions around the cation vacancy, further highlighting improvements over fixed electronegativity models.

QEq models with fixed electronegativities are unable to account for changes in the tendency for an atom to attract or repel electrons with changes in the bonding environments \cite{Choudhury79, ILeven19}.  In contrast, our ACE model has the capability to predict changes in atomic electronegativities based on different bonding environments, which leads to more realistic charge distributions.

\begin{figure}
    \centering
    \includegraphics[width=0.50\textwidth]{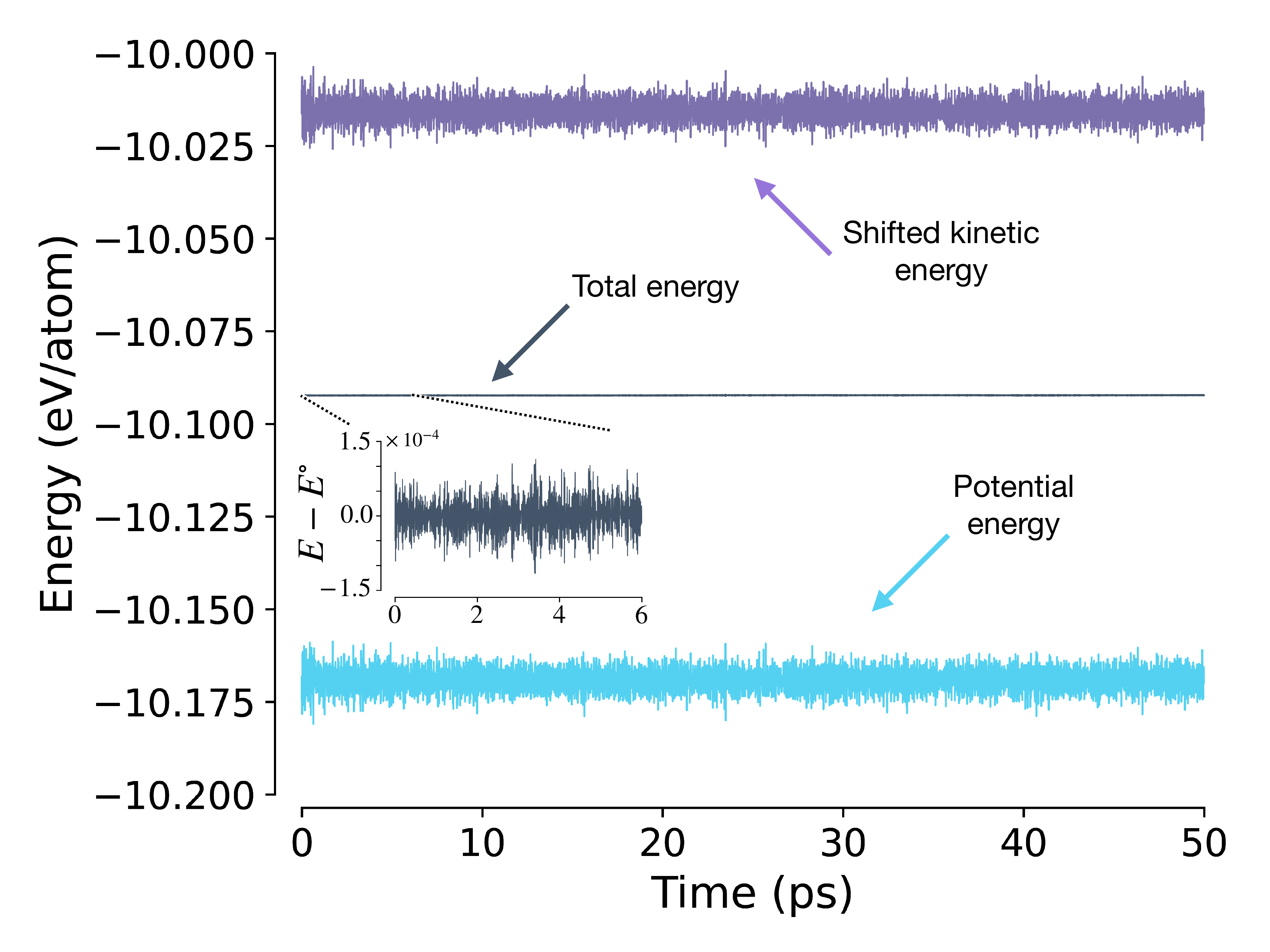}
    \caption{The total, potential, and shifted kinetic energy per atom is plotted for an ACE+XL-QEq shadow molecular dynamics simulation of a 323-atom UO$_2$ cell with a uranium vacancy.  The average statistical temperature was around 500 K in an NVE ensemble with an integration time step, $\delta t$ = 1.6 fs.  The inset shows the fluctuations of the total energy on a smaller scale.  The estimated energy drift was -0.094 $\mu$eV/atom $\cdot$ ps.}
    \label{fig:long_uo2}
\end{figure}

The effect of the local environment on the behavior of the electronegativities as a function of time during an ACE+XL-QEq simulation is shown in Fig.\ \ref{fig:eneg_plot}. These electronegativities are shown for uranium atoms (A), and oxygen atoms (B), both near and far from a vacancy center ($Va_U$).  Close to the vacancy we notice significant shifts and larger fluctuations.  Conversely, atoms positioned further away from the vacancy have electronegativity values closer to those in bulk UO$_2$.  This is in close agreement with our ground truth data generated from the SCC-DFTB theory.  For example, the SCC-DFTB simulations predict that oxygen has an average electronegativity (and standard deviation) of 3.5(0.2) V for an oxygen close to a Uranium vacancy and 3.1(0.1) V in the bulk part of the crystal, in close agreement with our ACE+XL-QEq simulation.  With a fixed electronegativity model, there is of course no fluctuation in the electronegativity, and with a fixed value of 3.3, averaged from all oxygen atoms, the partial charges would deviate significantly from the ground truth both near and far from the vacancy.  A QEq model with fixed electronegativities would only be adequate for bulk UO$_2$ simulation without any defects.  A significantly improved representation of atomic electronegativities is thus obtained with the ACE+XL-QEq scheme during the shadow molecular dynamics simulations. 
The ACE+XL-QEq scheme dynamically predicts changes in the atomic electronegativities that lead to a more realistic charge distribution in varied chemical environments.  This is achieved at a modest extra cost.  The computational cost of the variable ACE electronegativities and their force contributions in the ACE+XL-QEq shadow molecular dynamics simulations is about the same as for the charge-independent potential, $V_{\rm S}({\bf R})$, and its force contributions. 

The long-term stability is an important gauge of the quality of a molecular dynamics simulation.  The long-term stability of an ACE+XL-QEq shadow molecular dynamics simulation is demonstrated in Fig.\ \ref{fig:long_uo2}, where the total, potential, and kinetic energies per atom are plotted for a 323-atom UO$_2$ uranium vacancy system. The estimated drift in the energy over 50 ps of simulation time was only -0.094 $\mu$eV/atom $\cdot$ ps.  Similar stability was also observed in our other simulations of UO$_2$.

In the ACE+XL-QEq molecular dynamics simulations, partial charges, ${\bf n}(t)$, appear as extended dynamical variables that are propagated through a harmonic oscillator using the extended Lagrangian shadow molecular dynamics formalism outlined in earlier sections.  The accuracy and robustness of the ACE+XL-QEq simulations depend on how closely the ${\bf n}$-dependent ground state partial charges, ${\boldsymbol \eta}^{\rm min}[{\bf n}]$ in Eq.\ (\ref{Eta_Min}), determining the shadow Born-Oppenheimer potential, ${\cal U}_{\rm BO}({\bf R,n})$ in Eq.\ (\ref{U2Min}), follow the exact ground state charges, ${\boldsymbol \eta}^{\rm min}$ in Eq.\ (\ref{qmin}), of the regular exact Born-Oppenheimer potential, $U_{\rm BO}({\bf R})$ in Eq.\ (\ref{UMIN}). 
We can track the difference between  ${\boldsymbol \eta}^{\rm min}[{\bf n}]$ and ${\boldsymbol \eta}^{\rm min}$ by calculating the exact ground state charges, ${\boldsymbol \eta}^{\rm min}$, using a full matrix inversion in each time step. 

\begin{figure}
    \centering
    \includegraphics[width=0.45\textwidth]{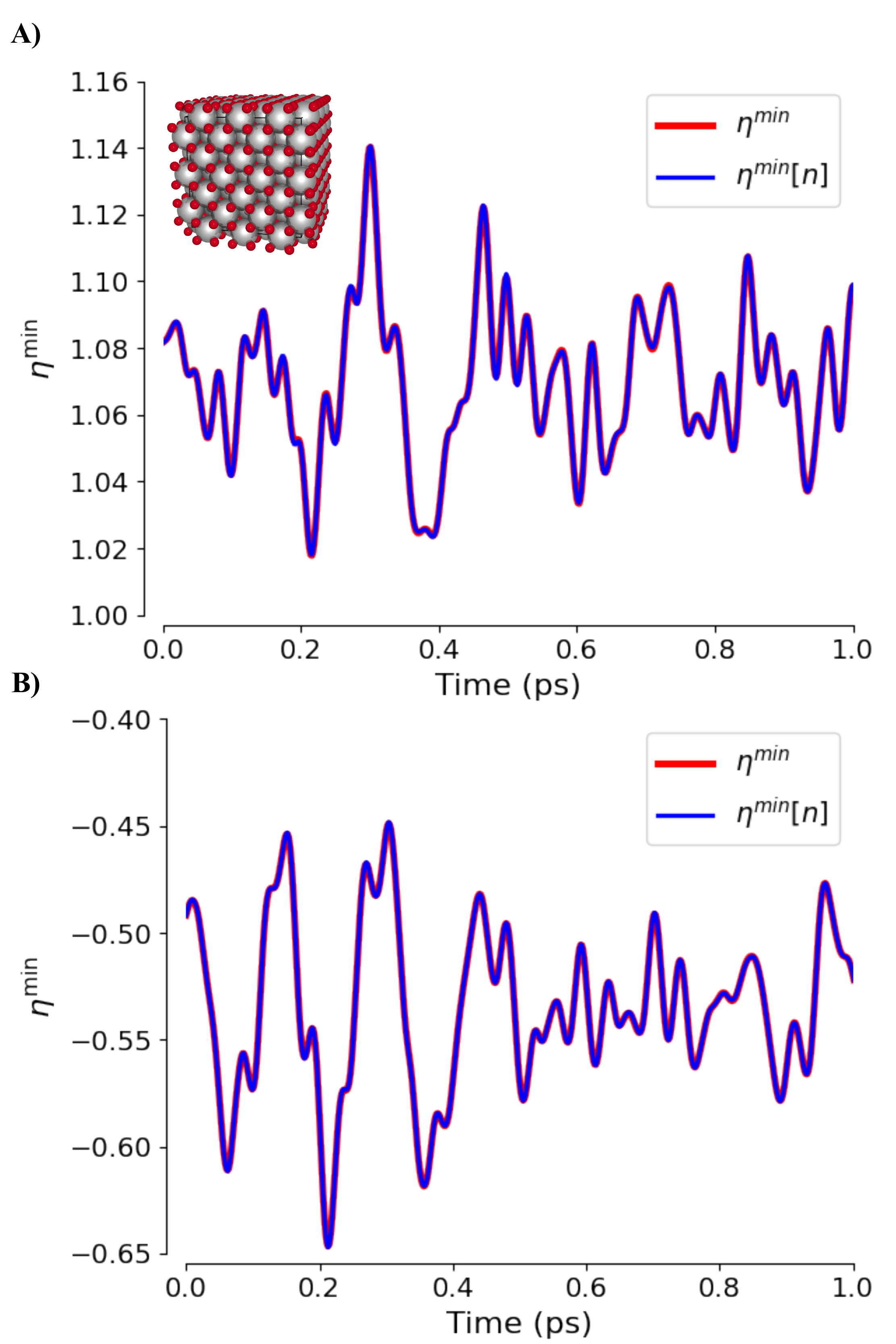}
    \caption{The exact (red) ground state charges, ${\boldsymbol \eta}^{\rm min}(t)$, of the regular Born-Oppenheimer potential and the ground state charges of the shadow Born-Oppenheimer potential, ${\boldsymbol \eta}^{\rm min}[{\bf n}](t)$, (blue) during the first 1 ps of an NVE simulation of UO$_2$ for a randomly selected uranium atom (A) and a randomly selected oxygen atom (B). The curves are virtually on top of each other for many different integration time steps (1.6 fs step plotted here).}
    \label{fig:full_charge_uo2}
\end{figure}

In Fig. \ref{fig:full_charge_uo2}, the exact ground state partial charges, ${\boldsymbol \eta}^{\rm min}$, of the regular Born-Oppenheimer potential and the corresponding ground state charges of the shadow Born-Oppenheimer potential, ${\boldsymbol \eta}^{\rm min}[{\bf n}]$, are compared for randomly selected U (upper panel A) and O (lower panel B) atoms from a 324-atom microcanonical (NVE) molecular dynamics simulation of UO$_2$ with a statistical temperature averaging about 1000 K.  The difference between the exact and XL-QEq ground state charges is very small, with an RMSE of $0.05 ~me$ on average over the simulation.

\begin{figure}[h!]
    \centering
    \includegraphics[width=0.45\textwidth]{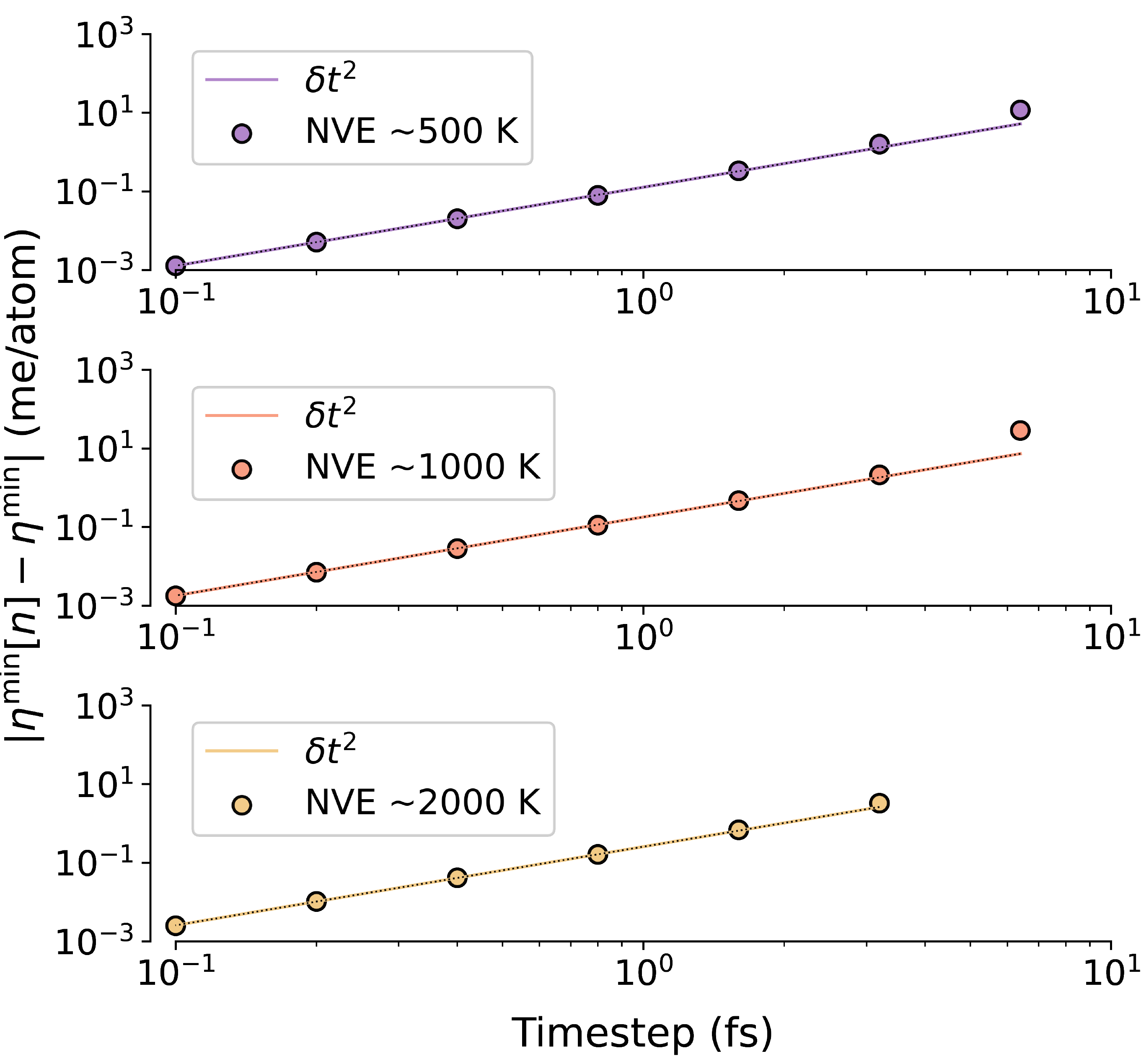}
    \caption{The average RMS difference per-atom between the exact charge and the dynamically propagated charge as a function of the time step for NVE simulations of the UO$_2$ system. The ideal $\delta t^2$ scaling is plotted for comparison.}
    \label{fig:eta_scale_U}
\end{figure}

The ground state charges of the shadow molecular dynamics are stable and continue to closely follow the exact ones also for longer simulation time scales.  The partial charges generated by the ACE+XL-QEq shadow molecular dynamics scheme are thus virtually identical to the partial charges of the ``exact'' regular Born-Oppenheimer simulation, but without having to solve all-to-all systems of equations.

In addition to the fluctuations in the total energy, the RMS difference between the exact charges (calculated through direct matrix inversion) and the dynamically propagated partial atomic charges, ($|{\boldsymbol \eta}^{\rm min}(t) -{\bf n}(t)|$), should also scale with the square of the time integration step, $\delta t^2$ \cite{ANiklasson21, ANiklasson21b}. This is an important property that shows how the ACE+XL-QEq shadow molecular dynamics becomes exact in the limit of continuous time.  The relation is demonstrated in Fig.~\ref{fig:eta_scale_U} for the UO$_2$ system during NVE simulations over a large range of average simulation temperatures.

\subsubsection{H$_2$O}

\begin{figure}
    \centering
    \includegraphics[width=0.50\textwidth]{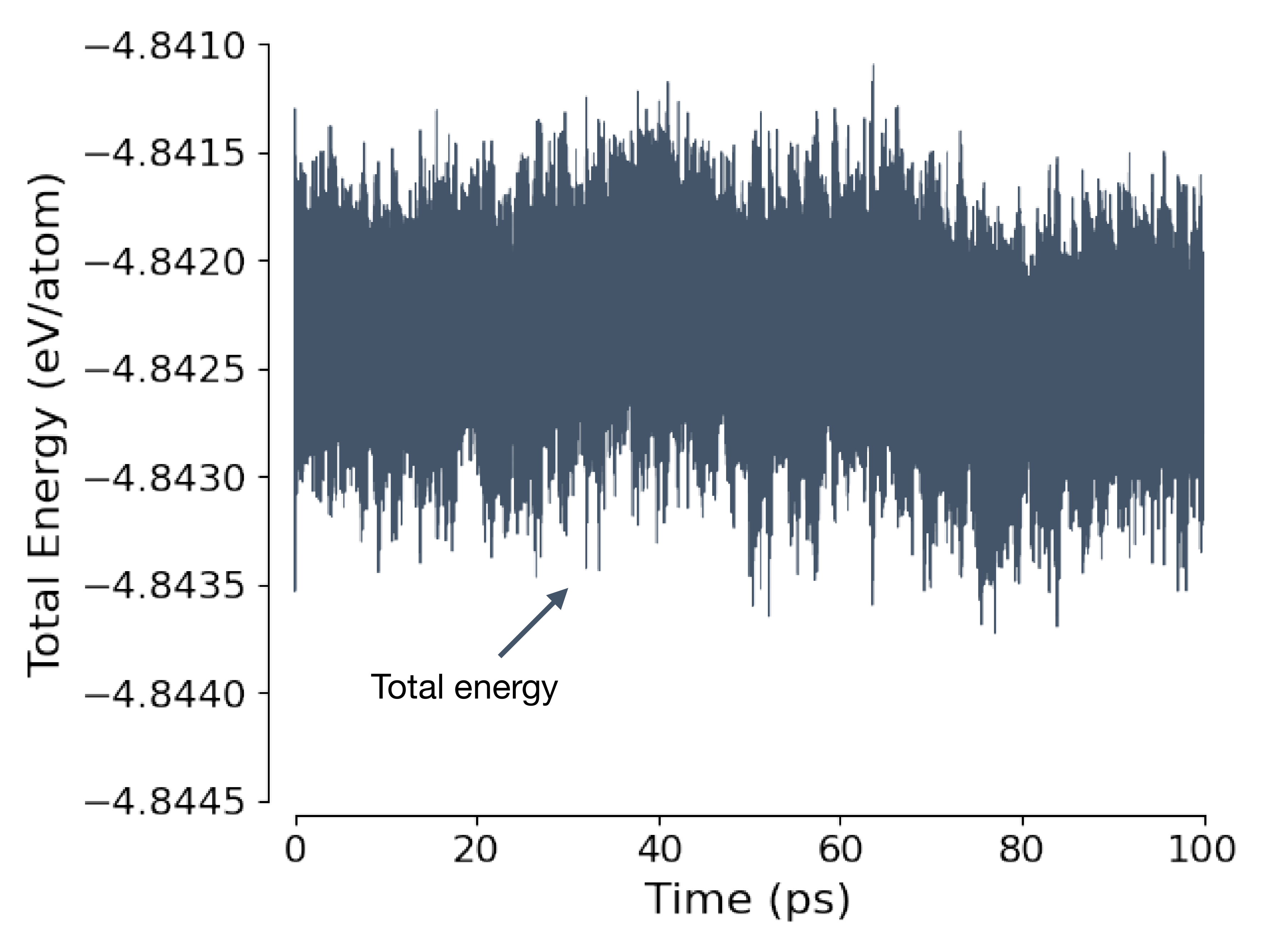}
    \caption{The fluctuations in the total energy for a 100 ps simulation of a 24 atom water cell with periodic boundary conditions. The average temperature was about 350 K, and the integration time step was 0.5 fs. A generalized first-level shadow Born-Oppenheimer potential was used \cite{Niklasson2023}. The estimated energy drift is about -2.3 $\mu$eV/atom $\cdot$ ps.}
    \label{fig:wat_long}
\end{figure}

The ACE+XL-QEq shadow molecular dynamics scheme also produces accurate and stable simulations for water. Fig.\ \ref{fig:wat_long} shows the total energy (potential + kinetic) during a simulation of a small periodic water system with 24 atoms,  and with a density of 1 g/ml near room temperature over a 100 ps simulation period. As with the UO$_2$ simulations, the preconditioner was updated every 500 steps, requiring a rank 2 update on average. Fairly small time integration steps were chosen to demonstrate the scaling of the accuracy, such as the error in the shadow Born-Oppenheimer potential, with respect to time step size, $\delta t$. The correct theoretical scaling is observed more clearly in the limit of shorter time steps. However, for the longer MD simulations of water, we are able to achieve stable integration with a time step of 0.5 fs, as shown in Fig.\ \ref{fig:wat_long}. In this case we use a generalized ``first-level'' approximation of the Born-Oppenheimer potential \cite{Niklasson2023} and the more accurate electronegativity model with the ACE coefficients trained to generate the partial atomic charges directly. The first-level approximation for the shadow Born-Oppenheimer potential may be obtained by performing one extra summation of the Coulomb potential and has an error that in the ideal case scales as $\delta t^8$ compared to the ``exact'' Born-Oppenheimer potential. More information about the modified first-level shadow shadow potential may be found in Ref. \cite{Niklasson2023}. The simulation provides excellent long-term stability with no systematic energy drift.

\begin{figure*}
        \centering
        \begin{subfigure}[b]{0.475\textwidth}
            \centering
            \includegraphics[width=\textwidth]{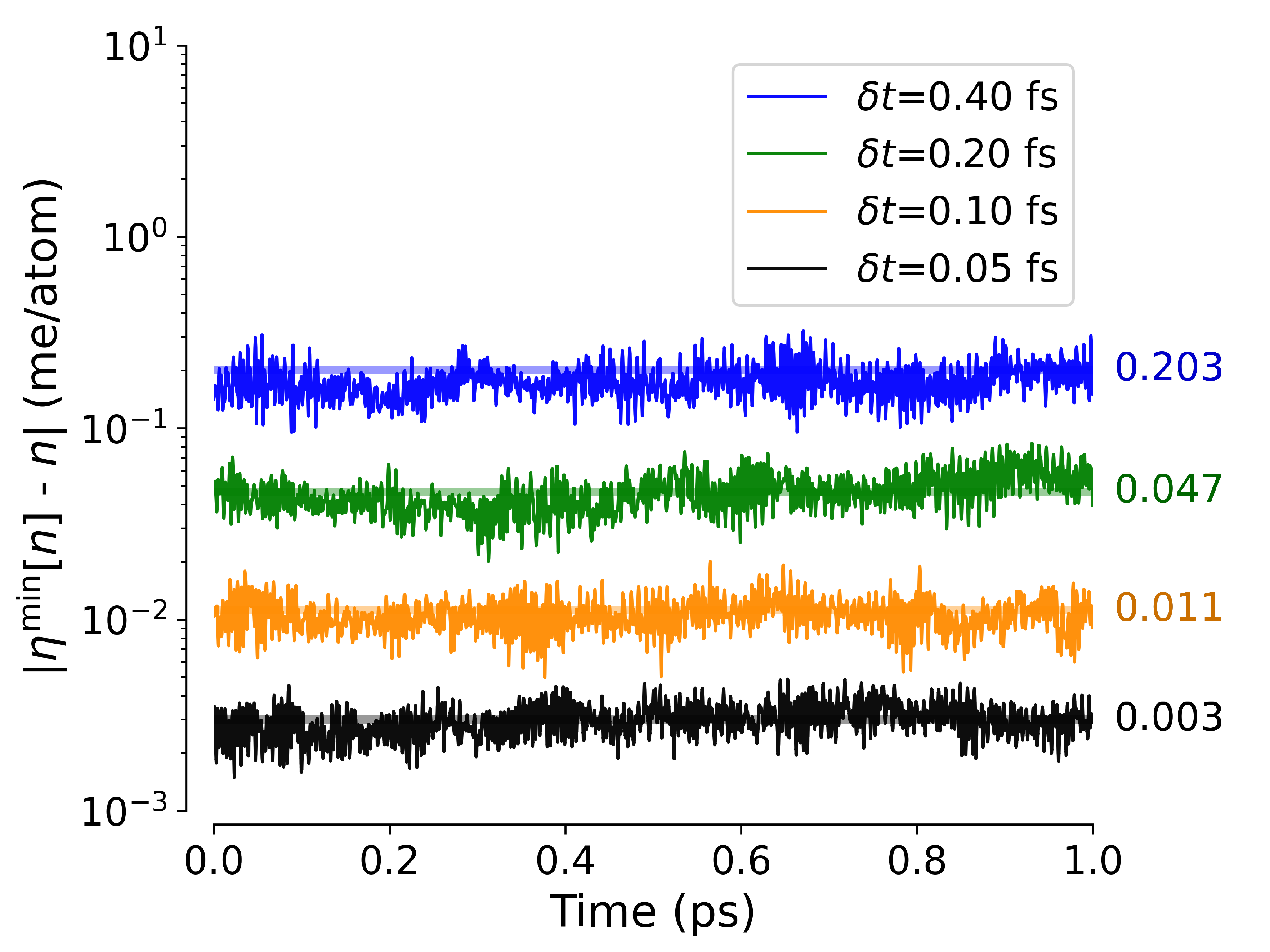}
            \caption[]%
            {{\small RMS of $|{\boldsymbol \eta}^{\rm{min}}[{\bf n}]-{\bf n}|$ with variable ACE electronegativities}}    
            \label{fig:chg_variable}
        \end{subfigure}
        \hfill
        \begin{subfigure}[b]{0.475\textwidth}
            \centering 
            \includegraphics[width=\textwidth]{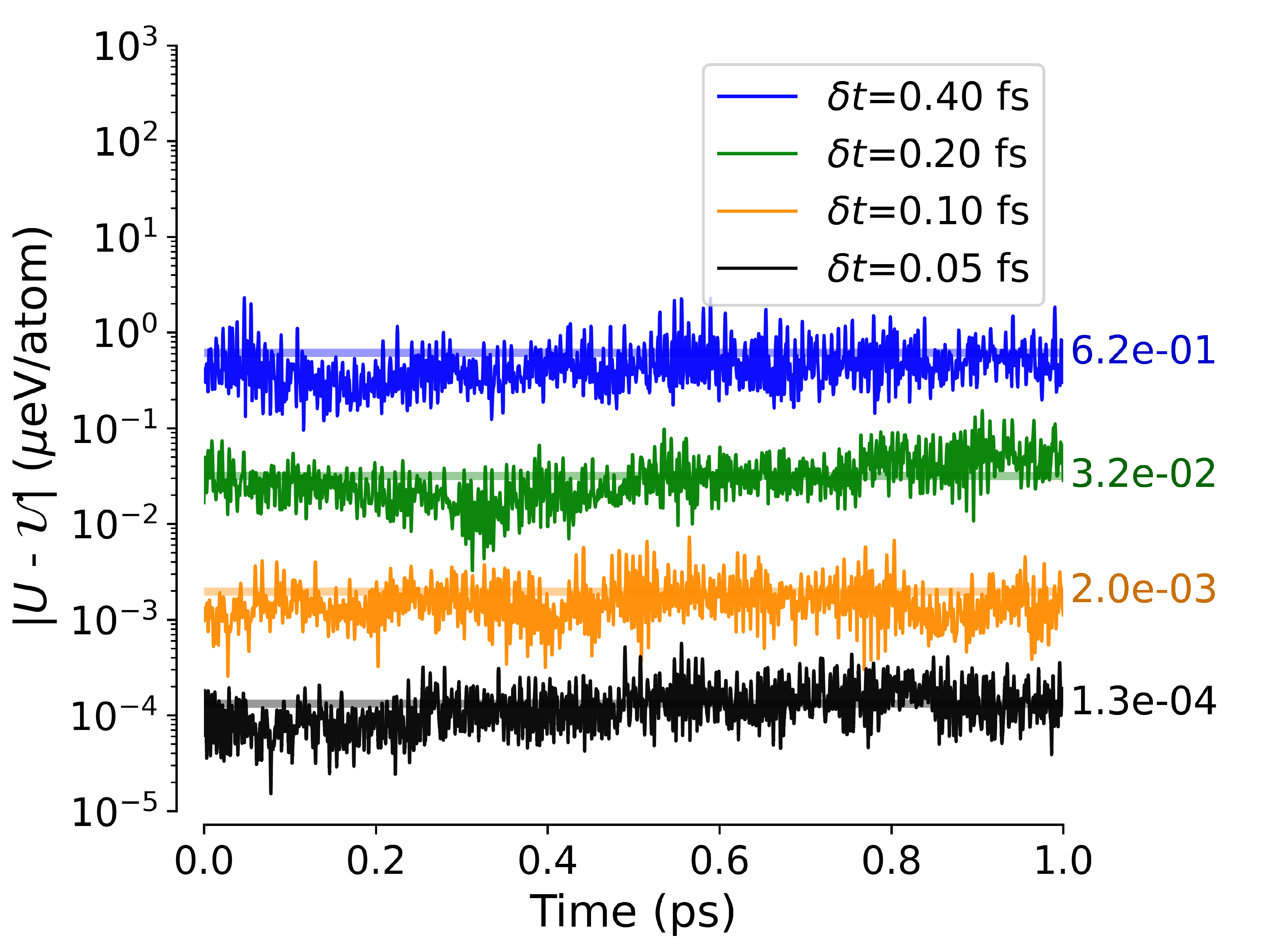}
            \caption[12]%
            {{\small  $|U_{\rm BO}({\bf R}) -{\cal U}_{\rm BO}({\bf R,n})|$ with variable ACE electronegativities}}    
            \label{fig:coul_variable}
        \end{subfigure}
        \vskip\baselineskip
        \begin{subfigure}[b]{0.475\textwidth}   
            \centering 
            \includegraphics[width=\textwidth]{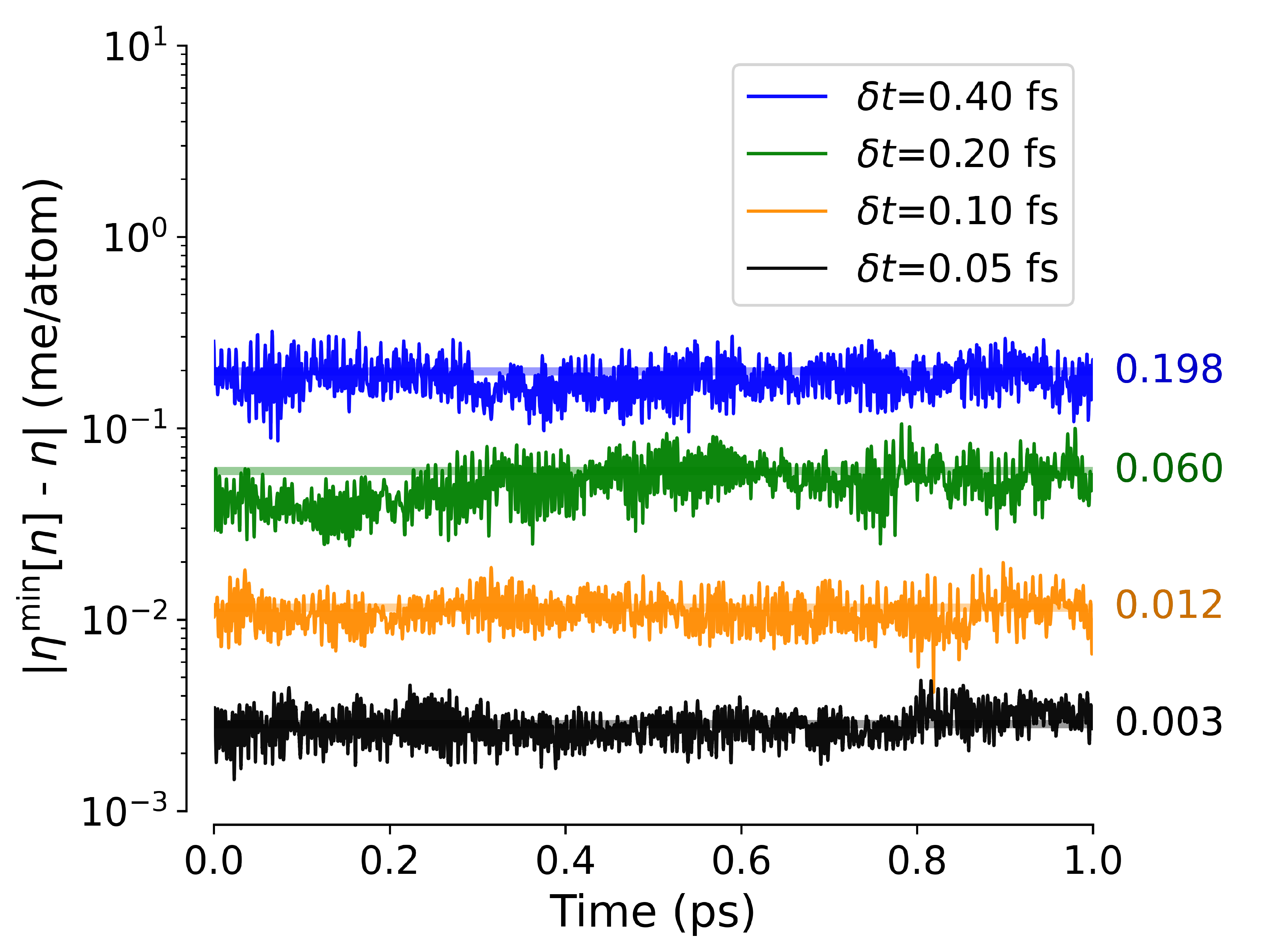}
            \caption[]%
            {{\small RMS of $|{\boldsymbol \eta}^{\rm{min}}[{\bf n}]-{\bf n}|$ with fixed electronegativities}}    
            \label{fig:chg_fixed}
        \end{subfigure}
        \hfill
        \begin{subfigure}[b]{0.475\textwidth}   
            \centering 
            \includegraphics[width=\textwidth]{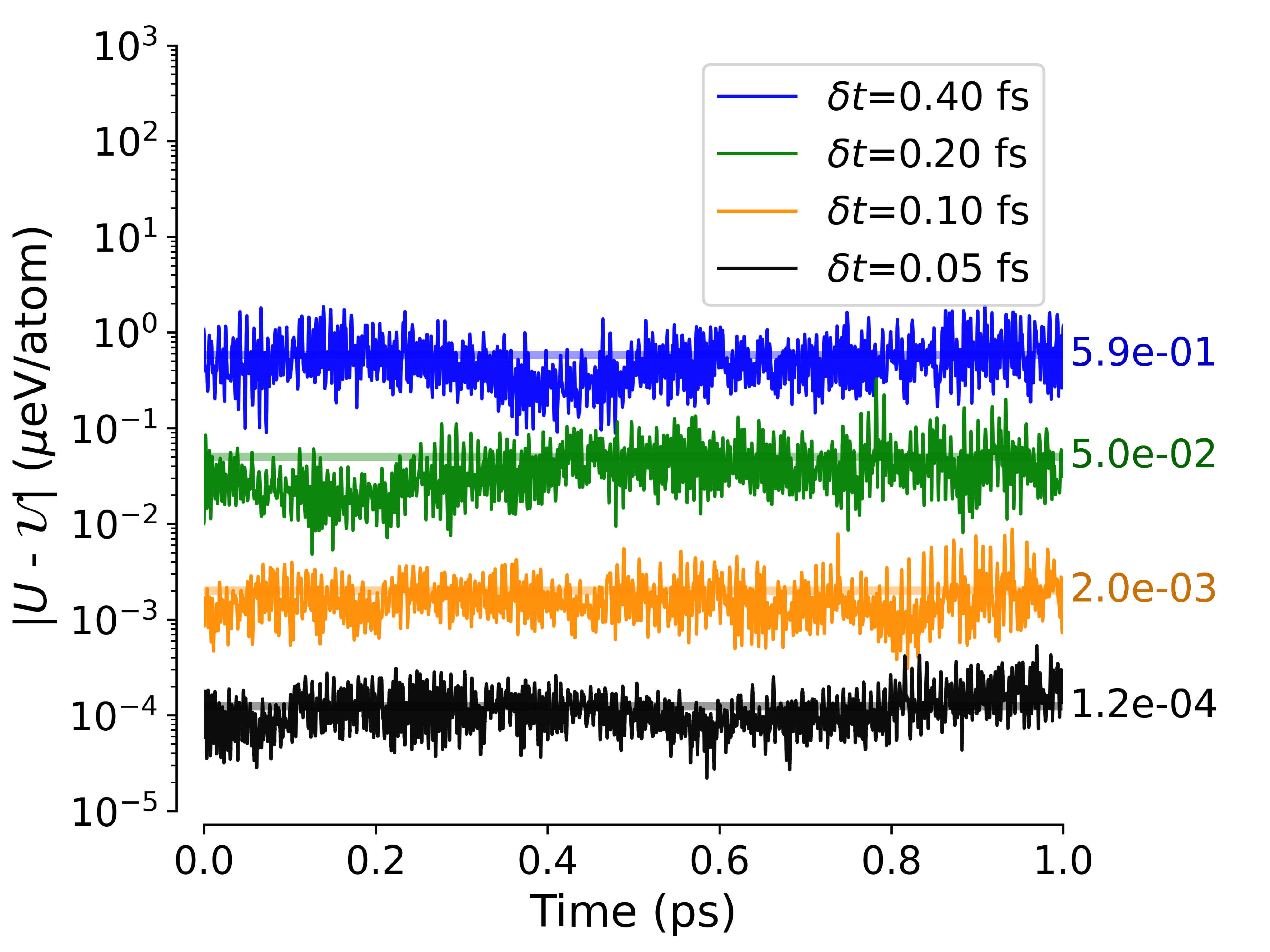}
            \caption[]%
            {{\small $|U_{\rm BO}({\bf R}) -{\cal U}_{\rm BO}({\bf R,n})|$ with fixed electronegativities}}    
            \label{fig:coul_fixed}
        \end{subfigure}
        \caption[  ]
        {\small Scaling of charge errors as well as Born-Oppenheimer potential energy errors with respect to integration time step for XL-QEq molecular dynamics simulations of a water system. Average values of errors over the 1 ps window of a water simulation are given on the right-hand side of each plot. Time steps are given in fs.} 
        \label{fig:varACE_vs_fixed}
\end{figure*}

To demonstrate the accuracy of the ACE+XL-QEq shadow molecular dynamics simulations for water we can compare the values of the shadow Born-Oppenheimer potential, ${\cal U}_{\rm BO}({\bf R,n})$, with the corresponding ``exact'' regular Born-Oppenheimer potential, $U_{\rm BO}({\bf R})$, generated by a matrix diagonalization. This comparison is illustrated on the right hand-side of Fig.\ \ref{fig:varACE_vs_fixed}. The simulations were performed for the same set of water system as in Fig.\ \ref{fig:wat_long}, but using the first electronegativity model, where the ACE coefficients are trained to reproduce the calculated atomic electronegativities. In contrast to the simulation in Fig.\ \ref{fig:wat_long} we also use the normal shadow Born-Oppenheimer potential, i.e.\ not the generalized first-level shadow potential  \cite{Niklasson2023}.  The difference, $\vert {\cal U}_{\rm BO}({\bf R,n}) - U_{\rm BO}({\bf R}) \vert$, are shown both for ACE+XL-QEq shadow molecular with variable ACE electronegativities (upper panel (b)) and for fixed electronegativities (lower panel (d)).  
The simulations in Fig.\ \ref{fig:varACE_vs_fixed} are performed for 4 different integration time steps, $\delta t$. On the right-hand side of each plot, we give the average value over a 1 ps window.
In both cases, (b) and (d), we find that the error in the sampling of the potential energy scales as $\delta t^4$, with a very small prefactor. The shadow Born-Oppenheimer potential is thus virtually identical to the exact fully optimized regular Born-Oppenheimer potential determined by equilibrated partial charges given through a direct matrix inversion. The same error behavior in the shadow potential has been previously observed also for quantum-mechanical molecular dynamics simulations based on XL-BOMD \cite{ANiklasson17,ANiklasson21b}. 

The $\delta t^4$-scaling of the potential sampling error can be explained from how the error in the Born-Oppenheimer potential energy depends on the residual error. The error in the Born-Oppenheimer potential  scales with the square of the RMS of the residual error function, i.e.\
\begin{equation}
    \vert {\cal U}_{\rm BO}({\bf R,n})  - U_{\rm BO}({\bf R}) \vert \propto \vert {\boldsymbol \eta}^{\rm min}[{\bf n}] - {\bf n} \vert^2,
\end{equation}
where the RMS of the residual function, $\vert {\boldsymbol \eta}^{\rm min}[{\bf n}] - {\bf n} \vert \propto \delta t^2$, scales with the square of the integration time step, $\delta t$. This quadratic scaling with $\delta t$ of the residual function is demonstrated in the two left panels of  Fig.\ \ref{fig:varACE_vs_fixed}. The upper panel shows the results for flexible ACE electronegativities and the lower shows the results for fixed electronegativities.  The right-hand side of each plot shows the average value over a 1 ps window, which increases approximately by a factor of four as the size of the integration time step is increased by a factor of two.

The ACE+XL-QEq shadow molecular dynamics simulations for liquid water show a close agreement with the ``exact'' regular Born-Oppenheimer molecular dynamics simulations, in the same way as they did for UO$_2$. Using flexible ACE electronegativities does not change the behavior.

\section{Summary and conclusions}

We have presented a shadow molecular dynamics scheme for flexible charge models, where the shadow Born-Oppenheimer potential is derived from a coarse-grained approximation of range-separated Hohenberg-Kohn density functional theory. The short-range parts of the energy terms are well-suited to be modeled by the linear ACE, which provides a highly efficient and systematically improvable alternative to many machine learning methods. To demonstrate the theory we used a second-order QEq model in combination with XL-BOMD, where the ACE was used to model the short-range charge-independent parts of the force field and the per-site electronegativities.
We used this ACE+XL-QEq scheme to model the dynamics of SCC-DFTB UO$_2$ and H$_2$O. The ACE+XL-QEq scheme provides a stable dynamics that avoids the explicit solution of an all-to-all system of equations that normally is required to determine the relaxed ground state charges prior to the force evaluations. This drastically reduces the computational cost, while the accuracy in the charges and in the sampling of the potential energy surface remains high.

The ACE-based electronegativity model for UO$_2$ yields accurate predictions of Coulomb energies per atom, with an average discrepancy of only 1 meV/atom compared to the SCC-DFTB reference data. This is attributed to the excellent fit of the electronegativity for this system, which exhibits a training error of 106 mV. This electronegativity model is capable of predicting differences in the electronegativity in a variety of bond environments including high-temperature structures and structures containing vacancies. This is in contrast to a typical fixed electronegativity model, which is not able to capture the effects of the shift and larger fluctuations in the electronegativity of an oxygen atom close to an Uranium vacancy in the UO$_2$ system.

The choice of training for the ACE electronegativity model affects the accuracy of the resulting model. Training the ACE parameters on the calculated ground truth electronegativities must include an additional unknown reference chemical potential parameter, $\mu^{\rm ref.}$, for each new atomic configuration. Training the ACE parameters directly on the net partial charges instead \cite{TWKo20}, avoids this additional step.
Both training methods provide good electronegativity models, but the latter approach seems to be provide a more accurate parameterization using the same amount of training data.
Including flexible electronegativities in the second-order charge equilibration model improves the modeling of the charge relaxations, as observed around the U vacancy in UO$_2$. The flexible electronegativities may also help avoid spuriously large transfer that can appear in fixed electronegativity models, e.g.\  where the electronic charge transfer from a single hydrogen surpasses 1 \textit{e}. However, the main purpose of our testbed systems, is not to provide new and more accurate simulation models for nuclear fuel and water, but only to demonstrate how our shadow molecular dynamics approach can be used to generate reliable and consistent trajectories compared to the corresponding ``exact'' Born-Oppenheimer dynamics. In this sense, the testbed systems are only intended to represent two archetypal model problems, one for solids and the other for liquids. 

The ACE+XL-QEq scheme demonstrates excellent long-term stability in the simulations with very small energy drifts both for UO$_2$ and H$_2$O. The shadow molecular dynamics follows closely the ``exact'' Born-Oppenheimer molecular dynamics both with respect to the 
potential energy surface and the charges.

The use of machine learned electronegativities in QEq models has been considered before \cite{TWKo20,Staacke_2022,Shao_2022}, but to the best of our knowledge no demonstration of molecular dynamics simulations have been performed using flexible electronegativities. The shadow molecular dynamics and ACEs for flexible charge models presented in this article represent an efficient approach to atomistic simulations and it opens the door to a broad range of new applications.

\section{Acknowledgements}

This work is supported by the U.S. Department of Energy Office of Basic Energy Sciences (FWP LANLE8AN) and by the U.S. Department of Energy through the Los Alamos National Laboratory. This research was also supported by the Exascale Computing Project (17-SC-20-SC), a collaborative effort of the U.S. Department of Energy Office of Science and the National Nuclear Security Administration. The training and usage of machine learned ACE models was supported by the U.S. Department
of Energy, Office of Fusion Energy Sciences (OFES) under Field Work Proposal
Number 20-023149. Discussions with Joshua Finkelstein, Danny Perez, Aidan Thompson, Linnea Andersson, and Enrique Martinez are gratefully acknowledged. Los Alamos National Laboratory is operated by Triad National Security, LLC, for the National Nuclear Security Administration of the U.S. Department of Energy Contract No. 892333218NCA000001. This article has been authored by an employee of National Technology and Engineering Solutions of Sandia, LLC under Contract No. DE-NA0003525 with the U.S. Department of Energy (DOE). Sandia National Laboratories is a multimission laboratory managed and operated by National Technology and Engineering Solutions of Sandia, LLC, a wholly owned subsidiary of Honeywell International Inc., for the U.S. Department of Energy’s National Nuclear Security Administration under contract DE-NA0003525. The employee owns all right, title and interest in and to the article and is solely responsible for its contents. The United States Government retains and the publisher, by accepting the article for publication, acknowledges that the United States Government retains a non-exclusive, paid-up, irrevocable, world-wide license to publish or reproduce the published form of this article or allow others to do so, for United States Government purposes. The DOE will provide public access to these results of federally sponsored research in accordance with the DOE Public Access Plan https://www.energy.gov/downloads/doe-public-access-plan. 

\section{Appendix}

In the equation of motion for the extended electronic degrees of freedom, Eq.\ (\ref{EOM_n}), we need to approximate the kernel, ${\bf K}$, and how it acts on the residual function, ${\bf f(n)} = {\boldsymbol \eta}^{\rm min}[{\bf n}] - {\bf n}$. This can be achieved using a preconditioned Krylov subspace approximation \cite{ANiklasson20}. First we rewrite the equations of motion in Eq.\ (\ref{EOM_n}) in an equivalent form,
\begin{equation}
{\bf \ddot n} = - \omega^2 \left({\bf K}_0{\bf J}\right)^{-1}{\bf K}_0\left({\boldsymbol \eta}^{\rm min}[{\bf n}] - {\bf n} \right),
\end{equation}
 where we have introduced a preconditioner ${\bf K}_0 \approx {\bf J}^{-1}$. If we use the notation
\begin{align}
{\bf f}_{{\bf v}_k}({\bf n}) = & \frac{ \partial {\bf f}({\bf n} + \lambda {\bf v}_k)}{\partial \lambda} \Big \vert_{\lambda = 0} = {\bf J} {\bf v}_k\\
{\widetilde {\bf f}}_{{\bf v}_k}({\bf n}) \equiv & ~{\bf K}_0 {\bf f}_{{\bf v}_k}({\bf n})\\
{\widetilde {\bf f}}{\bf (n)} =&~ {\bf K}_0  {\bf f(n)} = {\bf K}_0 \left({\boldsymbol \eta}^{\rm min}[{\bf n}] - {\bf n} \right),
\end{align}
it is possible to show that the preconditioned Jacobian, ${\bf K}_0{\bf J}$, can be approximated by a low-rank (rank $m$) approximation,
\begin{align}
{\bf K}_0{\bf J} \approx \sum_{kl}^m \left({\bf K}_0 {\bf f}_{{\bf v}_k} \right)L_{kl} {\bf v}_l^{\rm T} \equiv \sum_{kl}^m{\widetilde {\bf f}_{{\bf v}_k}} L_{kl} {\bf v}_l^{\rm T},
\end{align}
for some set of vectors $\{{\bf v}_k\}$ \cite{ANiklasson20}. Here ${\bf L} = {\bf O}^{-1}$, where $O_{ij} = {\bf v}_i^T{\bf v}_j$. The corresponding inverse is given by the pseudo inverse,
\begin{align}
{\widetilde {\bf K}} = \left({\bf K}_0{\bf J}\right)^{-1} \approx \sum_{kl} {\bf v}_k {\widetilde M_{kl}} {\widetilde {\bf f}_{{\bf v}_l}^T} ,
\end{align}
where ${\widetilde {\bf M}} = {\bf S}^{-1}$, with $S_{ij} = {\widetilde {\bf f}}_i^T{\widetilde {\bf f}}_j$ .

By choosing the vectors, $\{{\bf v}_k\}$, from an orthogonalized preconditioned Krylov subspace \cite{ANiklasson20},
\begin{align}
\left\{{\bf v}_k\right\} \in {\cal K}^\perp = {\rm span}^\perp \left\{ {\widetilde {\bf f}}({\bf n}), ({\bf K}_0 {\bf J}) {\widetilde {\bf f}}({\bf n}),({\bf K}_0 {\bf J})^2 {\widetilde {\bf f}}({\bf n}), \ldots  \right\},
\end{align}
we can rapidly reach a well-converged and accurate approximation for the integration of the electronic equation of motion, where
\begin{align} 
{\bf \ddot n} = - \omega^2 \left(\sum_{kl} {\bf v}_k {\widetilde M_{kl}} {\widetilde {\bf f}_{{\bf v}_l}}^T \right){\bf K}_0\left({\boldsymbol \eta}^{\rm min}[{\bf n}] - {\bf n} \right).
\end{align}
The Krylov subspace vectors are calculated using the expression for the Jacobian in Eq.\ (\ref{Jacobian}) and requires one Coulomb potential calculation per vector, ${\bf v}_k$.

There are many options for how the preconditioner can be constructed. For example, we may use a regularized exact kernel, ${\bf K}_0 = \left({\bf J - \epsilon I}\right)^{-1}$, calculated for the system at the initial time step, where $\epsilon$ is a small constant. The cost of this preconditioner can be high, but because the preconditioner can be reused over and over again during a molecular dynamics simulation, the overall computational overhead of constructing the preconditioner is in general very small. In all our demonstrations of molecular dynamics simulations the preconditioner used $\epsilon = 0$.

\normalem
\bibliographystyle{unsrt}
\bibliography{main.bib}

\begin{thebibliography}{100}

\bibitem{JBehler07}
J\"org Behler and Michele Parrinello.
\newblock Generalized neural-network representation of high-dimensional
  potential-energy surfaces.
\newblock {\em Phys. Rev. Lett.}, 98:146401, Apr 2007.

\bibitem{APBartok10}
Albert~P. Bart\'ok, Mike~C. Payne, Risi Kondor, and G\'abor Cs\'anyi.
\newblock Gaussian approximation potentials: The accuracy of quantum mechanics,
  without the electrons.
\newblock {\em Phys. Rev. Lett.}, 104:136403, Apr 2010.

\bibitem{MRupp12}
Matthias Rupp, Alexandre Tkatchenko, Klaus-Robert M\"uller, and O.~Anatole von
  Lilienfeld.
\newblock Fast and accurate modeling of molecular atomization energies with
  machine learning.
\newblock {\em Phys. Rev. Lett.}, 108:058301, Jan 2012.

\bibitem{APThompson15}
A~P Thompson, L~P Swiler, C~R Trott, S~M Foiles, and G~J Tucker.
\newblock Spectral neighbor analysis method for automated generation of
  quantum-accurate interatomic potentials.
\newblock {\em J. Comput. Phys.}, 285:316--330, March 2015.

\bibitem{Anatole15}
Raghunathan Ramakrishnan, Pavlo~O Dral, Matthias Rupp, and O~Anatole von
  Lilienfeld.
\newblock Big data meets quantum chemistry approximations: The
  {$\Delta$-Machine} learning approach.
\newblock {\em J. Chem. Theory Comput.}, 11(5):2087--2096, May 2015.

\bibitem{JBehler16}
J{\"o}rg Behler.
\newblock Perspective: Machine learning potentials for atomistic simulations.
\newblock {\em J. Chem. Phys.}, 145(17):170901, November 2016.

\bibitem{NLubbers18}
Nicholas Lubbers, Justin~S Smith, and Kipton Barros.
\newblock Hierarchical modeling of molecular energies using a deep neural
  network.
\newblock {\em J. Chem. Phys.}, 148(24):241715, June 2018.

\bibitem{FAFaber17}
Felix~A Faber, Luke Hutchison, Bing Huang, Justin Gilmer, Samuel~S Schoenholz,
  George~E Dahl, Oriol Vinyals, Steven Kearnes, Patrick~F Riley, and O~Anatole
  von Lilienfeld.
\newblock Prediction errors of molecular machine learning models lower than
  hybrid {DFT} error.
\newblock {\em J. Chem. Theory Comput.}, 13(11):5255--5264, November 2017.

\bibitem{JHan18}
Jiequn Han, Linfeng Zhang, Roberto Car, and Weinan E.
\newblock Deep potential: A general representation of a many-body potential
  energy surface.
\newblock {\em Commun. Comput. Phys.}, 23(3):629--639, 2018.

\bibitem{JSSmith19}
J.~S. Smith, B.~T. Nebgen, R.~Zubatyuk, N.~Lubbers, C.~Devereux, Kipton Barros,
  S.~Tretiak, O.~Isayev, and A.~E. Roitberg.
\newblock Approaching coupled cluster accuracy with a general-purpose neural
  network potential through transfer learning.
\newblock {\em Nature Comm.}, 8:13890, 2017.

\bibitem{Drautz19}
Ralf Drautz.
\newblock Atomic cluster expansion for accurate and transferable interatomic
  potentials.
\newblock {\em Phys. Rev. B}, 99:014104, Jan 2019.

\bibitem{TMiller20}
Zhuoran Qiao, Matthew Welborn, Animashree Anandkumar, Frederick~R Manby, and
  Thomas~F Miller, 3rd.
\newblock {OrbNet}: Deep learning for quantum chemistry using symmetry-adapted
  atomic-orbital features.
\newblock {\em J. Chem. Phys.}, 153(12):124111, September 2020.

\bibitem{Tkatchenko20}
Frank No{\'e}, Alexandre Tkatchenko, Klaus-Robert M{\"u}ller, and Cecilia
  Clementi.
\newblock Machine learning for molecular simulation.
\newblock {\em Annu. Rev. Phys. Chem.}, 71:361--390, April 2020.

\bibitem{Drautz21}
Yury Lysogorskiy, Cas van~der Oord, Anton Bochkarev, Sarath Menon, Matteo
  Rinaldi, Thomas Hammerschmidt, Matous Mrovec, Aidan Thompson, Gábor Csányi,
  Christoph Ortner, and Ralf Drautz.
\newblock Performant implementation of the atomic cluster expansion (pace):
  Application to copper and silicon, 2021.

\bibitem{AClark21}
Aurora~E Clark, Henry Adams, Rigoberto Hernandez, Anna~I Krylov, Anders M~N
  Niklasson, Sapna Sarupria, Yusu Wang, Stefan~M Wild, and Qian Yang.
\newblock The middle science: Traversing scale in complex {Many-Body} systems.
\newblock {\em ACS Cent Sci}, 7(8):1271--1287, August 2021.

\bibitem{WJMortier86}
Wilfried~J Mortier, Swapan~K Ghosh, and S~Shankar.
\newblock Electronegativity-equalization method for the calculation of atomic
  charges in molecules.
\newblock {\em J. Am. Chem. Soc.}, 108(15):4315--4320, July 1986.

\bibitem{AKRappe91}
Anthony~K. Rappe and William A.~Goddard III.
\newblock Charge equilibration for molecular dynamics simulations.
\newblock {\em J. Phys. Chem}, 95(8):3358--3363, 1991.

\bibitem{Baker15}
Christopher~M Baker.
\newblock Polarizable force fields for molecular dynamics simulations of
  biomolecules.
\newblock {\em Wiley Interdiscip. Rev. Comput. Mol. Sci.}, 5(2):241--254, March
  2015.

\bibitem{JZhifeng19}
Zhifeng Jing, Chengwen Liu, Sara~Y Cheng, Rui Qi, Brandon~D Walker, Jean-Philip
  Piquemal, and Pengyu Ren.
\newblock Polarizable force fields for biomolecular simulations: Recent
  advances and applications.
\newblock {\em Annu. Rev. Biophys.}, 48:371--394, May 2019.

\bibitem{SGoedecker15}
S.~Alireza Ghasemi, Albert Hofstetter, Santanu Saha, and Stefan Goedecker.
\newblock Interatomic potentials for ionic systems with density functional
  accuracy based on charge densities obtained by a neural network.
\newblock {\em Phys. Rev. B}, 92:045131, Jul 2015.

\bibitem{ANiklasson21b}
A.~M.~N. Niklasson.
\newblock Extended lagrangian born–oppenheimer molecular dynamics: from
  density functional theory to charge relaxation models.
\newblock {\em Eur. Phys. J. B}, 94:164, 2021.

\bibitem{TWKo20}
Tsz~Wai Ko, Jonas~A. Finkler, Stefan Goedecker, and Jörg Behler.
\newblock A fourth-generation high-dimensional neural network potential with
  accurate electrostatics including non-local charge transfer.
\newblock {\em Nature Comm.}, 12:398, 2021.

\bibitem{Staacke_2022}
Carsten~G Staacke, Simon Wengert, Christian Kunkel, G{\'a}bor Cs{\'a}nyi,
  Karsten Reuter, and Johannes~T Margraf.
\newblock Kernel charge equilibration: efficient and accurate prediction of
  molecular dipole moments with a machine-learning enhanced electron density
  model.
\newblock {\em Mach. Learn.: Sci. Technol.}, 3(1):015032, March 2022.

\bibitem{Shao_2022}
Yunqi Shao, Linnéa Andersson, Lisanne Knijff, and Chao Zhang.
\newblock Finite-field coupling via learning the charge response kernel.
\newblock {\em Electronic Structure}, 4(1):014012, mar 2022.

\bibitem{ANiklasson07}
Anders M.~N. Niklasson, C.~J. Tymczak, and M.~Challacombe.
\newblock Time-reversible ab initio molecular dynamics.
\newblock {\em J. Chem. Phys.}, 126:144103, 2007.

\bibitem{ANiklasson08}
Anders M.~N. Niklasson.
\newblock {Extended Born-Oppenheimer molecular dynamics}.
\newblock {\em Phys. Rev. Lett.}, 100:123004, 2008.

\bibitem{MCawkwell12}
M.~J. Cawkwell and A.~M.~N. Niklasson.
\newblock {Energy conserving, linear scaling Born-Oppenheimer molecular
  dynamics}.
\newblock {\em J. Chem. Phys.}, 137:134105, 2012.

\bibitem{JHutter12}
J.~Hutter.
\newblock Car parrinello molecular dynamics.
\newblock {\em WIREs Comput. Mol. Sci.}, 2:604, 2012.

\bibitem{LLin14}
L.~Lin, J.~Lu, and S.~Shao.
\newblock Analysis of time reversible born-oppenheimer molecular dynamics.
\newblock {\em Entropy}, 16:110, 2014.

\bibitem{PSouvatzis14}
P.~Souvatzis and A.~M.~N. Niklasson.
\newblock First principles molecular dynamics without self-consistent field
  optimization.
\newblock {\em J. Chem. Phys.}, 140:044117, 2014.

\bibitem{KNomura15}
K.~Nomura, P.~E. Small, R.~K. Kalia, A.~Nakano, and P.~Vashista.
\newblock An extended-lagrangian scheme for charge equilibration in reactive
  molecular dynamics simulations.
\newblock {\em Comput. Phys. Comm.}, 192:91, 2015.

\bibitem{AAlbaugh15}
A.~Albaugh, O.~Demardash, and T.~Head-Gordon.
\newblock An efficient and stable hybrid extended lagrangian/self-consistent
  field scheme for solving classical mutual induction.
\newblock {\em J. Chem. Phys.}, 143:174104, 2015.

\bibitem{ANiklasson17}
A.~M.~N. Niklasson.
\newblock {Next generation extended Lagrangian first principles molecular
  dynamics}.
\newblock {\em J. Chem. Phys.}, 147:054103, 2017.

\bibitem{AAlbaugh18}
Alex Albaugh, Teresa Head-Gordon, and Anders M~N Niklasson.
\newblock {Higher-Order} extended lagrangian {Born--Oppenheimer} molecular
  dynamics for classical polarizable models.
\newblock {\em J. Chem. Theory Comput.}, 14(2):499--511, February 2018.

\bibitem{ANiklasson21}
A.~M.~N. Niklasson.
\newblock Extended lagrangian born-oppenheimer molecular dynamics for
  orbital-free density functional theory and polarizable charge equilibration
  models.
\newblock {\em J. Chem. Phys.}, 154:0000, 2021.

\bibitem{Drautz20}
Ralf Drautz.
\newblock Atomic cluster expansion of scalar, vectorial, and tensorial
  properties including magnetism and charge transfer.
\newblock {\em Phys. Rev. B}, 102:024104, Jul 2020.

\bibitem{dusson_atomic_2022}
Geneviève Dusson, Markus Bachmayr, Gábor Csányi, Ralf Drautz, Simon Etter,
  Cas van~der Oord, and Christoph Ortner.
\newblock Atomic cluster expansion: {Completeness}, efficiency and stability.
\newblock {\em Journal of Computational Physics}, 454:110946, April 2022.

\bibitem{ACE_Carbon_22}
Minaam Qamar, Matous Mrovec, Yury Lysogorskiy, Anton Bochkarev, and Ralf
  Drautz.
\newblock Atomic cluster expansion for quantum-accurate large-scale simulations
  of carbon, 2022.

\bibitem{hohen}
P.~Hohenberg and W.~Kohn.
\newblock Inhomgenous electron gas.
\newblock {\em Phys. Rev.}, 136:B:864--B871, 1964.

\bibitem{KohnSham65}
W~Kohn and L.~J. Sham.
\newblock Self-consistent equations including exchange and correlation effects.
\newblock {\em Phys. Rev.}, 140(4A):1133, 1965.

\bibitem{RParr89}
R.~G. Parr and W.~Yang.
\newblock {\em Density-functional theory of atoms and molecules}.
\newblock Oxford University Press, Oxford, 1989.

\bibitem{RMDreizler90}
R.M. Dreizler and K.U. Gross.
\newblock {\em Density-functional theory}.
\newblock Springer Verlag, Berlin Heidelberg, 1990.

\bibitem{JPerdew92}
John~P. Perdew, J.~A. Chevary, S.~H. Vosko, Koblar~A. Jackson, Mark~R.
  Pederson, D.~J. Singh, and Carlos Fiolhais.
\newblock Atoms, molecules, solids, and surfaces: Applications of the
  generalized gradient approximation for exchange and correlation.
\newblock {\em Phys. Rev. B}, 46:6671--6687, Sep 1992.

\bibitem{WKohn99}
W.~Kohn.
\newblock Nobel lecture: Electronic structure of matter---wave functions and
  density functionals.
\newblock {\em Rev. Mod. Phys.}, 71:1253--1266, Oct 1999.

\bibitem{WHarrison80}
W.~A. Harrison.
\newblock {\em Electronic structure and the properties of solids: the physics
  of the chemical bond}.
\newblock Dover, New York, 1980.

\bibitem{MFoulkes89}
W.\ M.\~C. Foulkes and R.~Haydock.
\newblock Tight-binding models and density-functional theory.
\newblock {\em Phys. Rev. B}, 39:12520, 1989.

\bibitem{DPorezag95}
D.~Porezag, Th. Frauenheim, Th. K\"ohler, G.~Seifert, and R.~Kaschner.
\newblock Construction of tight-binding-like potentials on the basis of
  density-functional theory: Application to carbon.
\newblock {\em Phys. Rev. B}, 51:12947--12957, May 1995.

\bibitem{MElstner98}
M.~Elstner, D.~Poresag, G.~Jungnickel, J.~Elsner, M.~Haugk, T.~Frauenheim,
  S.~Suhai, and G.~Seifert.
\newblock Self-consistent-charge density-functional tight-binding method for
  simulations of complex materials properties.
\newblock {\em Phys. Rev. B}, 58:7260, 1998.

\bibitem{MFinnis98}
M.~W. Finnis, A.~T. Paxton, M.~Methfessel, and M.~van Schilfgarde.
\newblock Crystal structures of zirconia from first principles and
  self-consistent tight binding.
\newblock {\em Phys. Rev. Lett.}, 81:5149, 1998.

\bibitem{TFrauenheim00}
Th. Frauenheim, G.~Seifert, M.~Elstner, Z.~Hajnal, G.~Jungnickel, D.~Poresag,
  S.~Suhai, and R.~Scholz.
\newblock A self-consistent charge density-functional based tight-binding
  method for predictive materials simulations in physics, chemistry and
  biology.
\newblock {\em Phys. Stat. sol.}, 217:41, 2000.

\bibitem{PKoskinen09}
Pekka Koskinen and Ville M{\"a}kinen.
\newblock Density-functional tight-binding for beginners.
\newblock {\em Comput. Mater. Sci.}, 47(1):237--253, November 2009.

\bibitem{MGaus11}
M.~Gaus, Q.~Cui, and M.~Elstner.
\newblock Dftb3: Extension of the self-consistent-charge density-functional
  tight-binding method (scc-dftb).
\newblock {\em J, Chem. Theory Comput.}, 7:931, 2011.

\bibitem{BAradi15}
B.~Aradi, A.~M.~N. Niklasson, and Th. Frauenheim.
\newblock {Extended Lagrangian Density Functional Tight-Binding Molecular
  Dynamics for Molecules and Solids}.
\newblock {\em J. Chem. Theory Comput.}, 11:3357, 2015.

\bibitem{BHourahine20}
B~Hourahine, B~Aradi, V~Blum, F~Bonaf{\'e}, A~Buccheri, C~Camacho, C~Cevallos,
  M~Y Deshaye, T~Dumitric{\u a}, A~Dominguez, S~Ehlert, M~Elstner, T~van~der
  Heide, J~Hermann, S~Irle, J~J Kranz, C~K{\"o}hler, T~Kowalczyk, T~Kuba{\v r},
  I~S Lee, V~Lutsker, R~J Maurer, S~K Min, I~Mitchell, C~Negre, T~A Niehaus,
  A~M~N Niklasson, A~J Page, A~Pecchia, G~Penazzi, M~P Persson, J~{\v
  R}ez{\'a}{\v c}, C~G S{\'a}nchez, M~Sternberg, M~St{\"o}hr, F~Stuckenberg,
  A~Tkatchenko, V~W-Z Yu, and T~Frauenheim.
\newblock {DFTB+}, a software package for efficient approximate density
  functional theory based atomistic simulations.
\newblock {\em J. Chem. Phys.}, 152(12):124101, March 2020.

\bibitem{DMYork96}
D.~M. York and W.~Yang.
\newblock A chemical potential equalization method for molecular simulations.
\newblock {\em J.\ Chem.\ Phys.}, 104:159, 1996.

\bibitem{GTabacchi02}
G.~Tabacchi, C.~J. Mundy, J.~Hutter, and M.~Parrinello.
\newblock Classical polarizable force ﬁelds parametrized from ab initio
  calculations.
\newblock {\em J. Chem. Phys.}, 117:1416, 2002.

\bibitem{HYoshida90}
Haruo Yoshida.
\newblock Construction of higher order symplectic integrators.
\newblock {\em Phys. Lett. A}, 150(5):262--268, November 1990.

\bibitem{CGrebogi90}
C~Grebogi, S~M Hammel, J~A Yorke, and T~Sauer.
\newblock Shadowing of physical trajectories in chaotic dynamics: Containment
  and refinement.
\newblock {\em Phys. Rev. Lett.}, 65(13):1527--1530, September 1990.

\bibitem{SToxvaerd94}
S~Toxvaerd.
\newblock Hamiltonians for discrete dynamics.
\newblock {\em Phys. Rev. E Stat. Phys. Plasmas Fluids Relat. Interdiscip.
  Topics}, 50(3):2271--2274, September 1994.

\bibitem{GJason00}
Jason Gans and David Shalloway.
\newblock Shadow mass and the relationship between velocity and momentum in
  symplectic numerical integration.
\newblock {\em Phys. Rev. E}, 61:4587--4592, Apr 2000.

\bibitem{ShadowHamiltonian}
Stephen~D. Bond and Benedict~J. Leimkuhler.
\newblock {\em Molecular dynamics and the accuracy of numerically computed
  averages}.
\newblock Cambride University Press, United Kingdom, 2007.

\bibitem{SToxvaerd12}
S{\o}ren Toxvaerd, Ole~J Heilmann, and Jeppe~C Dyre.
\newblock Energy conservation in molecular dynamics simulations of classical
  systems.
\newblock {\em J. Chem. Phys.}, 136(22):224106, June 2012.

\bibitem{KDHammonds20}
K.~D. Hammonds and D.~M. Heyes.
\newblock Shadow hamiltonian in classical {NVE} molecular dynamics simulations:
  A path to long time stability.
\newblock {\em J. Chem. Phys.}, 152(2):024114, January 2020.

\bibitem{AAlbaugh17}
A.~Albaugh, Anders M.~N. Niklasson, and T.~Head-Gordon.
\newblock Accurate classical polarization solution with no self-consistent
  field iterations.
\newblock {\em J. Phys. Chem. Lett.}, 8:1714, 2017.

\bibitem{SPlimpton95}
Steve Plimpton.
\newblock Fast parallel algorithms for {Short-Range} molecular dynamics.
\newblock {\em J. Comput. Phys.}, 117(1):1--19, March 1995.

\bibitem{thompson_lammps_2022}
Aidan~P. Thompson, H.~Metin Aktulga, Richard Berger, Dan~S. Bolintineanu,
  W.~Michael Brown, Paul~S. Crozier, Pieter~J. in~'t Veld, Axel Kohlmeyer,
  Stan~G. Moore, Trung~Dac Nguyen, Ray Shan, Mark~J. Stevens, Julien Tranchida,
  Christian Trott, and Steven~J. Plimpton.
\newblock {LAMMPS} - a flexible simulation tool for particle-based materials
  modeling at the atomic, meso, and continuum scales.
\newblock {\em Comput. Phys. Commun.}, 271:108171, February 2022.

\bibitem{LATTE_LAMMPS}
fix latte command --- {LAMMPS} documentation.
\newblock \url{https://docs.lammps.org/fix_latte.html}.
\newblock Accessed: 2023-5-30.

\bibitem{LATTE_OLD}
E.~J. Sanville and et~al.
\newblock {\sc LATTE}.
\newblock \url{http://www.github.com/lanl/latte}, 2010.
\newblock \mbox{L}os Alamos National Laboratory (LA- CC-10004).

\bibitem{AKrishnapriyan17}
A.~Krishnapryian, P.~Yang, A.~M.~N. Niklasson, and M.~J. Cawkwell.
\newblock Numerical optimization of density functional tight binding models:
  Application to molecules containing carbon, hydrogen, nitrogen, and oxygen.
\newblock {\em J. Chem. Theory Comput.}, 13:6191, 2017.

\bibitem{Perriot2018-cg}
Romain Perriot, Christian F~A Negre, Shawn~D McGrane, and Marc~J Cawkwell.
\newblock Density functional tight binding calculations for the simulation of
  shocked nitromethane with {LATTE-LAMMPS}.
\newblock {\em AIP Conf. Proc.}, 1979(1):050014, July 2018.

\bibitem{rohskopf2023fitsnap}
A~Rohskopf, C~Sievers, N~Lubbers, Ma~Cusentino, J~Goff, J~Janssen, M~McCarthy,
  D~Montes~Oca de~Zapiain, S~Nikolov, K~Sargsyan, D~Sema, E~Sikorski,
  L~Williams, A~Thompson, and M~Wood.
\newblock {FitSNAP}: Atomistic machine learning with {LAMMPS}.
\newblock {\em J. Open Source Softw.}, 8(84):5118, April 2023.

\bibitem{Note1}
Here we chose to ignore self-interaction corrections, which may affect the
  locality of the short-range functional $F_S[\rho ]$.

\bibitem{Note2}
This also means that ${\protect \bf n}$ is close to the exact ground state of
  the regular Born-Oppenheimer potential, ${\protect \boldsymbol \eta
  }^{\protect \rm min}$.

\bibitem{HCAndersen80}
H~C Andersen.
\newblock Molecular dynamics simulations at constant pressure and/or
  temperature.
\newblock {\em J. Chem. Phys.}, 1980.

\bibitem{MParrinello80}
M~Parrinello and A~Rahman.
\newblock Crystal structure and pair potentials: A {Molecular-Dynamics} study.
\newblock {\em Phys. Rev. Lett.}, 45(14):1196--1199, October 1980.

\bibitem{SNose84}
Shuichi Nos{\'e}.
\newblock A unified formulation of the constant temperature molecular dynamics
  methods.
\newblock {\em J. Chem. Phys.}, 81(1):511--519, July 1984.

\bibitem{RCar85}
R.~Car and M~Parrinello.
\newblock Unified approach for molecular dynamics and density-functional
  theory.
\newblock {\em Phys. Rev. Lett.}, 55:2471, 1985.

\bibitem{DRemler90}
D.~K. Remler and P.~A. Madden.
\newblock Molecular dynamics without effective potentials via the
  car-parrinello approach.
\newblock {\em Mol.\ Phys.}, 70:921, 1990.

\bibitem{GPastore91}
G.~Pastore, E.~Smargassi, and F.~Buda.
\newblock Theory of ab initio molecular-dynamics calculations.
\newblock {\em Phys. Rev. A}, 44:6334, 1991.

\bibitem{FBornemann98}
F.~A. Bornemann and C.~Sch\"{u}tte.
\newblock A mathematical investigation of the car-parrinello method.
\newblock {\em Numerische Mathematik}, 78:359, 1998.

\bibitem{DMarx00}
D.~Marx and J.~Hutter.
\newblock {\em Modern Methods and Algorithms of Quantum Chemistry}.
\newblock ed. J. Grotendorst, John von Neumann Institute for Computing,
  J\"ulich, Germany, second edition, 2000.

\bibitem{MTuckerman02}
M.\~E.\ Tuckerman.
\newblock Ab initio molecular dynamics: basic concepts, current trends and
  novel applications.
\newblock {\em J. Phys.: Conden. Matter}, 14:1297, 2002.

\bibitem{GZerah92}
G.~Zerah, J.~J. Clerouin, and E.~L. Pollock.
\newblock {\em Phys. Rev. Lett.}, 69:446, 1992.

\bibitem{JClerouin92}
J.~J. Clerouin, G.~Zerah, and E.~L. Pollock.
\newblock {\em Phys. Rev. A}, 46:5130, 1992.

\bibitem{FLambert06}
F.~Lambert, J.~Clerouin, and S.~Mazevet.
\newblock {\em Eur. Phys. Lett.}, 75:681, 2006.

\bibitem{MSprik88}
M~Sprik and M~L Klein.
\newblock A polarizable model for water using distributed charge sites.
\newblock {\em J. Chem. Phys.}, 1988.

\bibitem{MSprik90}
M.~Sprik.
\newblock Computer simulation of the dynamics of induced polarization
  fluctuations in water.
\newblock {\em J. Chem. Phys.}, 95:2283--2291, 1990.

\bibitem{DVanBelle92}
D.~Van~Belle, M.~Froeyen, G.~Lippens, and S.~J. Wodak.
\newblock Molecular dynamics simulation of polarizable water by an extended
  lagrangian method.
\newblock {\em Mol. Phys.}, 77:239--266, 1992.

\bibitem{GLamoureaux03}
G.~Lamoureux and B.~T. Roux.
\newblock Modeling induced polarization with classical drude oscillators:
  Theory and molecular dynamics simulation algorithm.
\newblock {\em J. Chem. Phys.}, 119:3025--3039, 2003.

\bibitem{BHartke92}
B~Hartke and E~A Carter.
\newblock Ab initio molecular dynamics with correlated molecular wave
  functions: Generalized valence bond molecular dynamics and simulated
  annealing.
\newblock {\em J. Chem. Phys.}, 1992.

\bibitem{HBSchlegel01}
H.\~B. Schlegel, J.\~M. Millam, S.\~S. Iyengar, G.\~A. Voth, A.\~D. Daniels,
  G.E. Scusseria, and M.\~J. Frisch.
\newblock Ab initio molecular dynamics: Propagating the density matrix with
  gaussian orbitals.
\newblock {\em J. Chem. Phys.}, 114:9758, 2001.

\bibitem{SIyengar01}
S.\~S. Iyengar, H.\~B. Schlegel, J.\~M. Millam, G.\~A. Voth, G.E. Scusseria,
  and M.\~J. Frisch.
\newblock Ab initio molecular dynamics: Propagating the density matrix with
  gaussian orbitals. ii. generalizations based on mass-weighting, idempotency,
  energy conservation and choice of initial conditions.
\newblock {\em J. Chem. Phys.}, 115:10291, 2001.

\bibitem{JMHerbert04}
J.\~M. Herbert and M.~Head-Gordon.
\newblock {Curvy-steps approach to constraint-free extended-Lagrangian ab
  initio molecular dynamics, using atom-centered basis functions: Convergence
  toward Born-Oppenheimer trajectories}.
\newblock {\em J. Chem. Phys.}, 121:11542, 2004.

\bibitem{JLi16}
J.~Li, C.~Haycraft, and S.\~S. Iyengar.
\newblock {\em J. Chem. Theory Comput.}, 12:2493, 2016.

\bibitem{JHarris85}
J.~Harris.
\newblock Simplified method for calculating the energy of weakly interacting
  fragments.
\newblock {\em Phys. Rev. B}, 31:1770--1779, 1985.

\bibitem{ANiklasson14}
A.~M.~N. Niklasson and M.J. Cawkwell.
\newblock {Generalized extended Lagrangian Born-Oppenheimer molecular
  dynamics}.
\newblock {\em J. Chem. Phys.}, 141:164123, 2014.

\bibitem{Niklasson2023}
Anders Niklasson and Christian F~A Negre.
\newblock Shadow energy functionals and potentials in {Born--Oppenheimer}
  molecular dynamics.
\newblock {\em J. Chem. Phys.}, 158(15), 2023.

\bibitem{ACoretti20}
A.~Coretti, L.~Scalfi, C.~Bacon, B.~Rotenberg, R.~Vuilleumier, G.~Ciccotti,
  M.~Salanne, and S.~Bonella.
\newblock Mass-zero constrained molecular dynamics for electrode charges in
  simulations of electrochemical systems.
\newblock {\em The Journal of Chemical Physics}, 152(19):194701, 2020.

\bibitem{SBonella20}
Sara Bonella, Alessandro Coretti, Rodolphe Vuilleumier, and Giovanni Ciccotti.
\newblock Adiabatic motion and statistical mechanics via mass-zero constrained
  dynamics.
\newblock {\em Phys. Chem. Chem. Phys.}, 22:10775--10785, 2020.

\bibitem{SBonella21}
D.~D. Girardier, A.~Coretti, G.~Ciccotti, and S.~Bonella.
\newblock Mass-zero constrained dynamics and statistics for the shell model in
  magnetic field.
\newblock {\em Eur. Phys. J. B}, 94:158, 2021.

\bibitem{ANiklasson09}
A.~M.~N. Niklasson, Peter Steneteg, Anders Odell, Nicolas Bock, Matt
  Challacombe, C.~J. Tymczak, Erik Holmstrom, Guishan Zheng, and Valery Weber.
\newblock {Extended Lagrangian Born-Oppenheimer molecular dynamics with
  dissipation}.
\newblock {\em J. Chem. Phys.}, 130:214109, 2009.

\bibitem{PSteneteg10}
P.~Steneteg, I.~A. Abrikosov, V.~Weber, and A.~M.~N. Niklasson.
\newblock {Wave function extended Lagrangian Born-Oppenheimer molecular
  dynamics}.
\newblock {\em Phys. Rev. B}, 82:075110, 2010.

\bibitem{GZheng11}
G.~Zheng, A.~M.~N. Niklasson, and M.~Karplus.
\newblock {Lagrangian formulation with dissipation of Born-Oppenheimer
  molecular dynamics using the density-functional tight-binding method}.
\newblock {\em J. Chem. Phys.}, 135:044122, 2011.

\bibitem{AOdell09}
A.~Odell, A.~Delin, B.~Johansson, N.~Bock, M.~Challacombe, and A.~M.~N.
  Niklasson.
\newblock {Higher-order symplectic integration in Born-Oppenheimer molecular
  dynamics}.
\newblock {\em J. Chem. Phys.}, 131:244106, 2009.

\bibitem{AOdell11}
A.~Odell, A.~Delin, B.~Johansson, M.~J. Cawkwell, and A.~M.~N. Niklasson.
\newblock Geometric integration in born-oppenheimer molecular dynamics.
\newblock {\em J. Chem. Phys.}, 135:224105, 2011.

\bibitem{VVitale17}
V.~Vitale, J.~Dziezic, A.~Albaugh, A.M.N. Niklasson, T.~J. Head-Gordon, and C-K
  Skylaris.
\newblock {Performance of extended Lagrangian schemes for molecular dynamics
  simulations with classical polarizable force fields and density functional
  theory}.
\newblock {\em J. Chem. Phys.}, 12:124115, 2017.

\bibitem{ILeven19}
I.~Leven and T.~Head-Gordon.
\newblock Inertial extended-lagrangian scheme for solving charge equilibration
  models.
\newblock {\em Phys. Chem. Chem. Phys.}, 21(34):18652--18659, 2019.

\bibitem{ANiklasson20}
A.~M.~N. Niklasson.
\newblock Extended lagrangian born-oppenheimer molecular dynamics using a
  krylov subspace approximation.
\newblock {\em J. Chem. Phys.}, 152:104103, 2020.

\bibitem{MAllen90}
M.~Allen and D.~Tildesley.
\newblock {\em Computer Simulation of Liquids}.
\newblock Oxford Science, London, 1990.

\bibitem{Gabor10}
Albert~P. Bart\'ok, Mike~C. Payne, Risi Kondor, and G\'abor Cs\'anyi.
\newblock Gaussian approximation potentials: The accuracy of quantum mechanics,
  without the electrons.
\newblock {\em Phys. Rev. Lett.}, 104:136403, Apr 2010.

\bibitem{Faber17}
F.~A. Faber, L.~Hutchison, B.~Huang, J.~Gilmer, S.~S. Schoenholz, G.~E. Dahl,
  O.~Vinyals, S.~Kearnes, P.~F. Riley, and O.~Anatole von Lilienfeld.
\newblock Machine learning prediction errors better than dft accuracy.
\newblock 2017.

\bibitem{KSchutt17}
K.~T. Schütt, F.~Arbabzadah, S.~Chmiela, K.~R. M\"{u}ller, and A.~Tkatchenko.
\newblock Quantum-chemical insights from deep tensor neural networks.
\newblock {\em Nature Comm.}, 8:13890, 2017.

\bibitem{Parsaeifard_2021}
Behnam Parsaeifard, Deb~Sankar De, Anders~S Christensen, Felix~A Faber, Emir
  Kocer, Sandip De, Jörg Behler, O~Anatole von Lilienfeld, and Stefan
  Goedecker.
\newblock An assessment of the structural resolution of various fingerprints
  commonly used in machine learning.
\newblock {\em Machine Learning: Science and Technology}, 2(1):015018, apr
  2021.

\bibitem{MLReview22}
N.~Fedik, R.~Zubatyuk, M.~Kulichenko, N.~Lubbers, J.~S. Smith, B.~Nebgen,
  R.~Messerly, Y-W. Li, A.~I. Boldyrev, K.~Barros, O.~Isayev, and S.~Tretiak.
\newblock Extending machine learning beyond interatomic potentials for
  predicting molecular properties.
\newblock {\em Nature Reviews Chemitry}, 6:653, 2022.

\bibitem{goff_permutation-adapted_2022}
James~M. Goff, Charles Sievers, Mitchell~A. Wood, and Aidan~P. Thompson.
\newblock Permutation-adapted complete and independent basis for atomic cluster
  expansion descriptors, August 2022.
\newblock arXiv:2208.01756 [cond-mat].

\bibitem{wipf_new_2007}
David Wipf and Srikantan Nagarajan.
\newblock A {New} {View} of {Automatic} {Relevance} {Determination}.
\newblock In {\em Advances in {Neural} {Information} {Processing} {Systems}},
  volume~20. Curran Associates, Inc., 2007.

\bibitem{ziegler1985stopping}
James~F Ziegler and Jochen~P Biersack.
\newblock {\em The stopping and range of ions in matter}.
\newblock Springer, 1985.

\bibitem{VIAnisimov91}
V.~I. Anisimov, J.~Zaanen, and O.~K. Andersen.
\newblock Band theory and mott insulators: Hubbard u instead of stoner i.
\newblock {\em Phys. Rev. B}, 44:943, 1991.

\bibitem{VIAnisimov97}
V.~I. Anisimov, F.~Aryasetiawan, and A.~I. Lichtenstein.
\newblock First-principles calculations of the electronic structure and spectra
  of strongly correlated systems: the lda + u method.
\newblock {\em J. Phys.: Condens. Matter}, 9:767, 1997.

\bibitem{AILichtenstein95}
A.~I. Lichtenstein, V.~I. Anisimov, and J.~Zaane.
\newblock Density-functional theory and strong interactions: Orbital ordering
  in mott-hubbard insulators.
\newblock {\em Phys. Rev. B}, 52:5467, 1995.

\bibitem{SLDudarev98}
S.~L. Dudarev, G.~A. Botton, S.~Y. Savrasov, C.~J. Humphreys, and A.~P. Sutton.
\newblock Electron-energy-loss spectra and the structural stability of nickel
  oxide: An lsda u study.
\newblock {\em Phys. Rev. B}, 57:1505, 1998.

\bibitem{HIdriss10}
H.~Idriss.
\newblock Surface reactions of uranium oxide powder, thin films and single
  crystals.
\newblock {\em Surface Science Reports}, 65(3):67--109, 2010.

\bibitem{HHeming13}
Heming He, David~A. Andersson, David~D. Allred, and Kirk~D. Rector.
\newblock Determination of the insulation gap of uranium oxides by
  spectroscopic ellipsometry and density functional theory.
\newblock {\em The Journal of Physical Chemistry C}, 117(32):16540--16551,
  2013.

\bibitem{HKulik15}
H.~Kulik.
\newblock Treating electron over-delocalization with the dft+u method.
\newblock {\em J. CHem. Phys.}, 142:240901, 2015.

\bibitem{zuo_performance_2020}
Yunxing Zuo, Chi Chen, Xiangguo Li, Zhi Deng, Yiming Chen, Jörg Behler, Gábor
  Csányi, Alexander~V. Shapeev, Aidan~P. Thompson, Mitchell~A. Wood, and
  Shyue~Ping Ong.
\newblock Performance and {Cost} {Assessment} of {Machine} {Learning}
  {Interatomic} {Potentials}.
\newblock {\em The Journal of Physical Chemistry A}, 124(4):731--745, January
  2020.
\newblock Publisher: American Chemical Society.

\bibitem{scikit-learn}
F.~Pedregosa, G.~Varoquaux, A.~Gramfort, V.~Michel, B.~Thirion, O.~Grisel,
  M.~Blondel, P.~Prettenhofer, R.~Weiss, V.~Dubourg, J.~Vanderplas, A.~Passos,
  D.~Cournapeau, M.~Brucher, M.~Perrot, and E.~Duchesnay.
\newblock Scikit-learn: Machine learning in {P}ython.
\newblock {\em Journal of Machine Learning Research}, 12:2825--2830, 2011.

\bibitem{Choudhury79}
M.~H. Choudhury and D.~B. Pearson.
\newblock Spectral properties of many-electron atomic hamiltonians and the
  method of configuration interaction .1. proof of convergence of the
  configuration interaction method.
\newblock {\em Journal of Mathematical Physics}, 20(4):752--756, 1979.

\end{thebibliography}

\end{document}